\documentclass[runningheads,envcountsame]{llncs}

\usepackage{amsmath,amssymb,amsfonts,mathtools}
\usepackage{hyperref}
\usepackage[pdftex,dvipsnames]{xcolor}
\usepackage{enumerate}
\usepackage{enumitem}
\usepackage{graphicx}
\usepackage{xargs}             
\usepackage{xspace}
\usepackage[normalem]{ulem}
\usepackage{booktabs} 
\usepackage{underscore} 
\usepackage{tabularx}
\usepackage{adjustbox}
\usepackage{wrapfig}
\usepackage{multirow, multicol}
\usepackage{booktabs}
\usepackage{multirow}
\usepackage{amsmath}
\usepackage{microtype}  
\usepackage{amssymb}
\newcolumntype{C}[1]{>{\centering\arraybackslash}m{#1}}

\usepackage{verified-badges}

\usepackage{colortbl} 

\errorcontextlines\maxdimen

\makeatletter
\newcommand*{\algrule}[1][\algorithmicindent]{\makebox[#1][l]{\hspace*{.5em}\vrule height .75\baselineskip depth .25\baselineskip}}%

\newcount\ALG@printindent@tempcnta
\def\ALG@printindent{%
    \ifnum \theALG@nested>0
        \ifx\ALG@text\ALG@x@notext
            \addvspace{-3pt}
        \else
            \unskip
            \ALG@printindent@tempcnta=1
            \loop
                \algrule[\csname ALG@ind@\the\ALG@printindent@tempcnta\endcsname]%
                \advance \ALG@printindent@tempcnta 1
            \ifnum \ALG@printindent@tempcnta<\numexpr\theALG@nested+1\relax
            \repeat
        \fi
    \fi
    }%
\usepackage{etoolbox}
\makeatletter
\patchcmd{\ALG@doentity}{\noindent\hskip\ALG@tlm}%
  {\ALG@printindent\noindent\hskip\ALG@tlm}{}{}
\makeatother

\usepackage{float}
\usepackage{placeins}
\usepackage{algorithm}
\usepackage{algpseudocode}
\usepackage{amsmath}
\usepackage{lineno}
\usepackage{hyperref} 
\usepackage{caption}

\algtext*{EndIf}
\algtext*{EndFor}
\algtext*{EndWhile}
\algtext*{EndProcedure}

\makeatletter
\renewenvironment{proof}{\par\noindent\textit{Proof.}\ }{\hfill$\square$\par}
\makeatother

\usepackage[T1]{fontenc}
\usepackage{amsmath}
\usepackage{amssymb}
\usepackage{graphicx}
\usepackage{multirow}
\usepackage{booktabs}
\usepackage{array}

\usepackage{booktabs}   

\usepackage[]{todonotes}

\newcommandx{\lt}[2][1=]{\todo[inline,
  linecolor=RawSienna,backgroundcolor=Salmon!40,bordercolor=RawSienna,#1]{\bfred{(Landon) #2}}}  

\newcommandx{\bh}[2][1=]{\todo[inline, linecolor=Cerulean, 
backgroundcolor=Cerulean!25, bordercolor=Cerulean]{\bfblue{(Beckey) #2}}}
\newcommandx{\bhzz}[2][1=]{\todo[inline, linecolor=Cerulean, 
  backgroundcolor=Cerulean!25, bordercolor=Cerulean]{\bfblue{(Zhen
      \textrightarrow\ Beckey) #2}}}

\newcommandx{\zz}[2][1=]
{\todo[inline,linecolor=LimeGreen,backgroundcolor=SpringGreen!40,bordercolor=LimeGreen,#1]{\bfgreen{(Zhen) #2}}}
\newcommandx{\zzbh}[2][1=]
{\todo[inline,linecolor=LimeGreen,backgroundcolor=SpringGreen!40,bordercolor=LimeGreen,#1]{\bfgreen{(Beckey \textrightarrow\ Zhen) #2}}}

\newcommandx{\tom}[2][1=]{\todo[inline,linecolor=Plum,backgroundcolor=Plum!25,bordercolor=Plum,#1]{\textcolor{Plum}{\textbf{(Thomas) #2}}}}
\newcommandx{\tombh}[2][1=]{\todo[inline,linecolor=Plum,backgroundcolor=Plum!25,bordercolor=Plum,#1]{\textcolor{Plum}{\textbf{(Thomas) From Beckey: #2}}}}

\newcommandx{\all}[2][1=]{\todo[inline,
linecolor=Bittersweet,backgroundcolor=Bittersweet!40,bordercolor=Bittersweet,#1]{\bfyellow{(All) #2}}}

\newcommandx{\rc}[2][1=]{\todo[inline,
linecolor=Bittersweet,backgroundcolor=Bittersweet!40,bordercolor=Bittersweet,#1]{\bfyellow{(Reviewer) #2}}}

\newcommand{\bfred}[1]{\textsf{\bfseries\color{Maroon}#1}\xspace}
\newcommand{\bfblue}[1]{\textsf{\bfseries\color{BlueViolet}#1}\xspace}
\newcommand{\bfgreen}[1]{\textsf{\bfseries\color{OliveGreen}#1}\xspace}
\newcommand{\bfyellow}[1]{\textsf{\bfseries\color{Bittersweet}#1}\xspace}

\usepackage{hmabs}        

\begin{document}

\title{Harmonic-Mean Aggregation of Continuous-Time Markov Chains with Reachability Error Bounds}
\titlerunning{Harmonic-Mean Aggregation of Continuous-Time Markov Chains}
\author{Bingqing Hu\inst{1}\orcidID{0009-0006-5933-6141} \and
Landon Taylor\inst{1}\orcidID{0000-0002-4071-3625} \and
Zhen Zhang\inst{1}\orcidID{0000-0002-8269-9489}}
\authorrunning{B. Hu et al.}
\institute{Department of Electrical and Computer Engineering, Utah State University, Logan, UT 84321, USA\\
\email{beckey.hu@usu.edu, landon.jeffrey.taylor@usu.edu, zhen.zhang@usu.edu}}
\maketitle

\begin{abstract}
Abstracting a continuous-time Markov chain by grouping its states into blocks raises a
fundamental question: which rates should the abstract chain use when states within a block
have different exit rates and transition probabilities? We study harmonic-mean
aggregation, which matches the mean waiting time and next-block distribution of a local
experiment that uniformly selects a state within the block and observes its next jump.
Equivalently, the abstract transition rates are averages of the concrete rates with
inverse-exit-rate weights. Since the aggregation does not preserve reachability
probabilities in general, we derive theoretical error bounds for time-bounded reachability. We
decompose the error into a blockwise mean and centered-within-block deviations, and show
that the rate mismatch has zero weighted mean. Exploiting this cancellation and
differences between abstract reachability values, we obtain bounds valid at every time
horizon through a linear comparison system with two quantities per block, without
computing concrete reachability values. All results are formalized and proven in Lean~4.

\end{abstract}
\section{Introduction}\label{sec:intro}

\emph{Continuous-time Markov chains} (CTMCs) serve as a fundamental modeling formalism for representing stochastic behaviors in systems ranging from biochemical reaction networks~\cite{Heath2008,Madsen2015,Buecherl2021,Friedenberg2022} to distributed communication protocols\cite{Kwiatkowska2001,Kwiatkowska2002}. Their capability to incorporate both timing and probabilistic dynamics renders them particularly effective for analyzing systems subject to real-time constraints and stochastic behaviors. However, formal verification of CTMCs is often intractable due to state explosion problem, where the number of reachable states increases exponentially with the number of system variables and interactions.

\emph{Probabilistic model checking} (PMC) verifies quantitative properties of probabilistic systems~\cite{Kwiatkowska2007,BaierKatoen2008}. A central task for CTMCs is \emph{time-bounded reachability}: computing the probability of reaching a goal within a given time horizon. Despite algorithmic advances, numerical transient analysis can require substantial computation over the reachable state space~\cite{Hensel2022}. Large or unbounded models therefore motivate approximate verification approaches with computable error bounds. This need is particularly relevant in domains such as synthetic biology and network security, where models capture stochastic dynamics over high-dimensional state spaces. State-space abstraction addresses this challenge by grouping concrete states into blocks. Constructing a smaller CTMC then requires a principled choice of transition rates that accounts for the different waiting times and destinations within each block.

This paper establishes theoretical foundations for \emph{harmonic-mean} (HM) aggregation, where abstract rates are derived from a local experiment that uniformly selects a concrete state within a block of states and observes its next concrete transition. Matching the mean waiting time yields a harmonic-mean event rate, while matching the destination-block distribution determines the abstract transition rates. Equivalently, the rates are weighted averages in which each state is weighted by its mean residence time. The uniform selection provides a reference distribution for this construction. To quantify the resulting reachability approximation, we introduce a \emph{value-aware error bound}. The analysis separates weighted block-mean error from centered within-block variation and exploits the cancellation of the weighted rate defect. A coupled linear system with two quantities per block bounds both components using concrete generator coefficients and differences between abstract reachability values. The resulting certificate reflects the selected goal and can be evaluated using the abstract reachability solution alone, together with the precomputed coefficients.

\paragraph{Contributions.} The main contributions of this paper are:
\begin{enumerate}
\item[(i)] Define HM aggregation, establish its local preservation (mean waiting time and destination law), and prove exactness on singleton partitions and under ordinary lumpability (Props.~\ref{prop:preserve},~\ref{prop:structure}).
\item[(ii)] Derive \emph{exact} equations for reachability mean error and intra-block variation (Lemma~\ref{lem:error-eq}) via an HM-weighted generator identity (Lemma~\ref{lem:hm-identity}), proving exactness for singletons and lumpable chains (Cor.~\ref{cor:exact}).
\item[(iii)] We prove a block-level, value-aware reachability error envelope for \emph{anytime} horizon and concrete initial state (Theorem~\ref{thm:envelope}), using centered and pinned support estimates.

\item[(iv)] We sketch an algorithm evaluating the bound on the abstract state space alone (Sect.~\ref{sec:err-eval}): its time-slab scheme makes the coefficients
piecewise constant, so each slab costs one matrix exponential
(Lemma~\ref{lem:eval}).

\item[(v)] We formalize and prove all results in Lean~4 with Mathlib.\footnote{Anonymous
repository: \url{https://anonymous.4open.science/r/hm-mean-9D42/lean/README.md}.} The formalization is supplementary and does
not replace the proofs in this paper.
\end{enumerate}

\paragraph{Automated proofs and AI use.}
Marked theorems and definitions are formalized in Lean~4~\cite{lean4} via Mathlib, 
with complete source code available\footnotemark[\value{footnote}]. 
Generative AI tools (Claude Code and Codex) assisted with language refinement, manuscript organization, exploratory discussions, 
and writing Lean formalizations. All Lean proofs are verified by the kernel, and all mathematical arguments were independently 
verified by the authors, who take full responsibility for the content.


\paragraph{Related work.}
Ordinary lumpability~\cite{Buchholz1994} yields exact aggregation, which HM
aggregation recovers; bounded aggregation~\cite{FranceschinisMuntz1994} bounds
quasi-lumpable chains. The $(\varepsilon,\delta)$-bisimulation for CTMCs of
Spork et al.~\cite{Spork2025} gives time- and reward-bounded reachability
guarantees for chains that are already close; we instead construct the
aggregate CTMC from a given predicate partition and certify its reachability
values.
Adaptive aggregation~\cite{AbateCAV2015,Ceska2021} and the formal error bounds
of Michel and Siegle~\cite{MichelSiegle2025} bound the error of aggregated
transient distributions; the latter admit weighted aggregations, which include
HM aggregation. Michel~\cite{Michel2025Wasserstein} extends them to
Wasserstein distance; Salamati et al.~\cite{SalamatiSM2020} derive
Lyapunov-based reachability bounds.
We instead give the HM rates \emph{local interpretation} and bound goal-dependent
reachability values directly, propagating block-mean error and within-block
spread jointly through a $2M$-dimensional comparison system.

Interval and imprecise abstractions bound CTMC behavior conservatively:
uniformisation with interval transition probabilities~\cite{KatoenKLW2007},
imprecise CTMCs on lumped state spaces~\cite{ErreygersDeBock2019}, and
lumpability under uncertain rates~\cite{CardelliEtAl2021}.
Population-interval abstractions~\cite{Ceska2019} combine interval
abstraction of populations with transition acceleration for order-of-magnitude
analysis of reaction networks. We average over the states in each block and
give an absolute error envelope for one ordinary CTMC.
Probabilistic CEGAR~\cite{Hermanns2008} targets discrete-time models;
fast adaptive uniformisation~\cite{fau} and sliding-window
abstraction~\cite{henzinger2009sliding} work on the fly or over time
intervals, whereas HM aggregation fixes one aggregate CTMC whose reachability
error is certified from the abstract solution alone.

\section{Preliminaries}\label{sec:preliminaries}

\paragraph{Continuous-time Markov chains.}\label{sec:ctmc}

A CTMC over atomic propositions $\AP$ is a tuple $\C = (\Sset, s_0, \R, L)$, where $\Sset$ is a finite state set, $s_0 \in \Sset$ is the initial state, $L \colon \Sset \to 2^{\AP}$ labels states, and $\R \colon \Sset \times \Sset \to \mathbb R_{\ge0}$ gives transition rates with no self-loops ($\R(s,s)=0$). The exit rate of $s$ is $E(s) = \sum_{s' \in \Sset} \R(s,s')$.
If $E(s) > 0$, state residence time $\tau_s \sim \Exp(E(s))$ and transition destination $\nu_s$ (where $\PP(\nu_s = s') = \R(s,s') / E(s)$) are independent. If $E(s) = 0$, $s$ is absorbing.
The generator $Q \in \mathbb R^{\Sset \times \Sset}$ has off-diagonal entries $Q_{ss'} = \R(s,s')$ ($s' \ne s$) and diagonal entries $Q_{ss} = -E(s)$, yielding zero row sums and non-positive diagonals.

\paragraph{Predicates, blocks, and abstract states.}\label{sec:predicates}

States are considered to be valuations of variables
$x_1,\ldots,x_n$ with finite integer domains $D_1,\ldots,D_n$, so that
$\Sset\subseteq D_1\times\cdots\times D_n$ (not every valuation need
be reachable). Abstraction is defined
by partitioning the domain of each variable into ranges.

\begin{figure}[bth]
 \centering
\scalebox{0.85}{%
\begin{tikzpicture}[x=0.44cm,y=0.5cm]
 \fill[hmblue!14]   (2.6,1.5) rectangle (4.5,4.6);
 \fill[hmorange!14] (7.5,1.5) rectangle (9.5,4.6);
 \foreach \i in {0,...,9}{\foreach \j in {0,...,6}{\fill[black!70] (\i,\j) circle (0.08);}}
 \foreach \i/\j in {0/6,1/5,9/0,8/1,9/6,7/6}{\fill[white] (\i,\j) circle (0.16);}
 \foreach \c in {2.6,4.5,7.5}{\draw[dashed,semithick,gray] (\c,-0.5) -- (\c,6.5);}
 \foreach \c in {1.5,4.6}{\draw[dashed,semithick,gray] (-0.5,\c) -- (9.5,\c);}
 \draw[semithick] (-0.5,-0.5) rectangle (9.5,6.5);
 \fill[hmblue]   (3,3) circle (0.17);
 \fill[hmorange] (8,3) circle (0.17);
 \node[font=\scriptsize,hmblue,anchor=south]   at (3,3.2) {$s$};
 \node[font=\scriptsize,hmorange,anchor=south] at (8,3.2) {$s'$};
 \draw[-{Stealth[length=1.6mm]},semithick,bend left=30] (3.12,3.08) to (7.88,3.08);
 \node[font=\scriptsize,hmblue]   at (3.55,4.33) {$B_k$};
 \node[font=\scriptsize,hmorange] at (8.5,4.33) {$B_\ell$};
 \foreach \x/\t in {1.05/1,3.55/2,6.0/3,8.5/4}{\node[font=\scriptsize] at (\x,7.1) {$I^1_\t$};}
 \foreach \y/\t in {0.5/1,3.05/2,5.55/3}{\node[font=\scriptsize,anchor=east] at (-0.7,\y) {$I^2_\t$};}
 \node[font=\footnotesize] at (4.5,8.3) {$\pi_1$ over $x_1$};
 \node[font=\footnotesize,rotate=90] at (-2.5,3) {$\pi_2$ over $x_2$};
 \node[font=\footnotesize,below] at (4.5,-0.6) {concrete states $\Sset$};

 \draw[-{Stealth[length=2.2mm]},very thick,hmgray] (10.1,3) -- (12.1,3);
 \node[font=\footnotesize,above] at (11.1,3.1) {$\alpha$};

 \begin{scope}[xshift=5.6cm]
  \foreach \c in {2.6,4.5,7.5}{\draw[dashed,gray!70] (\c,-0.5) -- (\c,6.5);}
  \foreach \c in {1.5,4.6}{\draw[dashed,gray!70] (-0.5,\c) -- (9.5,\c);}
  \draw[semithick] (-0.5,-0.5) rectangle (9.5,6.5);
  \foreach \x in {1.05,3.55,6.0,8.5}{\foreach \y in {0.5,3.05,5.55}{
    \node[circle,draw=black!55,fill=black!8,semithick,minimum size=4.6mm,inner sep=0pt] at (\x,\y) {};}}
  \node[circle,draw=hmblue,fill=hmblue!14,semithick,minimum size=5.6mm,inner sep=0pt,font=\scriptsize,text=hmblue]
       (sk) at (3.55,3.05) {$\hat s_k$};
  \node[circle,draw=hmorange,fill=hmorange!14,semithick,minimum size=5.6mm,inner sep=0pt,font=\scriptsize,text=hmorange]
       (sl) at (8.5,3.05) {$\hat s_\ell$};
  \draw[-{Stealth[length=1.6mm]},semithick] (sk) to[bend left=38] node[above=-1pt,font=\scriptsize] {$\Rhat$} (sl);
  \node[font=\footnotesize,below] at (4.5,-0.6) {abstract states $\Shat$};
 \end{scope}
\end{tikzpicture}%
}
 \caption{From predicates to abstract states. Left: $\pi_1$ over $x_1$ and $\pi_2$ over $x_2$ cut the
 domains into intervals $I^1_j$, $I^2_j$ of arbitrary width, and the nonempty boxes are the blocks
 (dots are states of $\Sset$). Right: $\alpha$ maps all of $B_k$ to $\hat s_k$. The transition
 $s\to s'$ between the non-adjacent blocks $B_k$ and $B_\ell$ contributes to $\rkl{k}{\ell}{s}$,
 and $\Chat$ has the single rate $\Rhat(\hat s_k,\hat s_\ell)$.}
 \label{fig:intervals}
\end{figure}
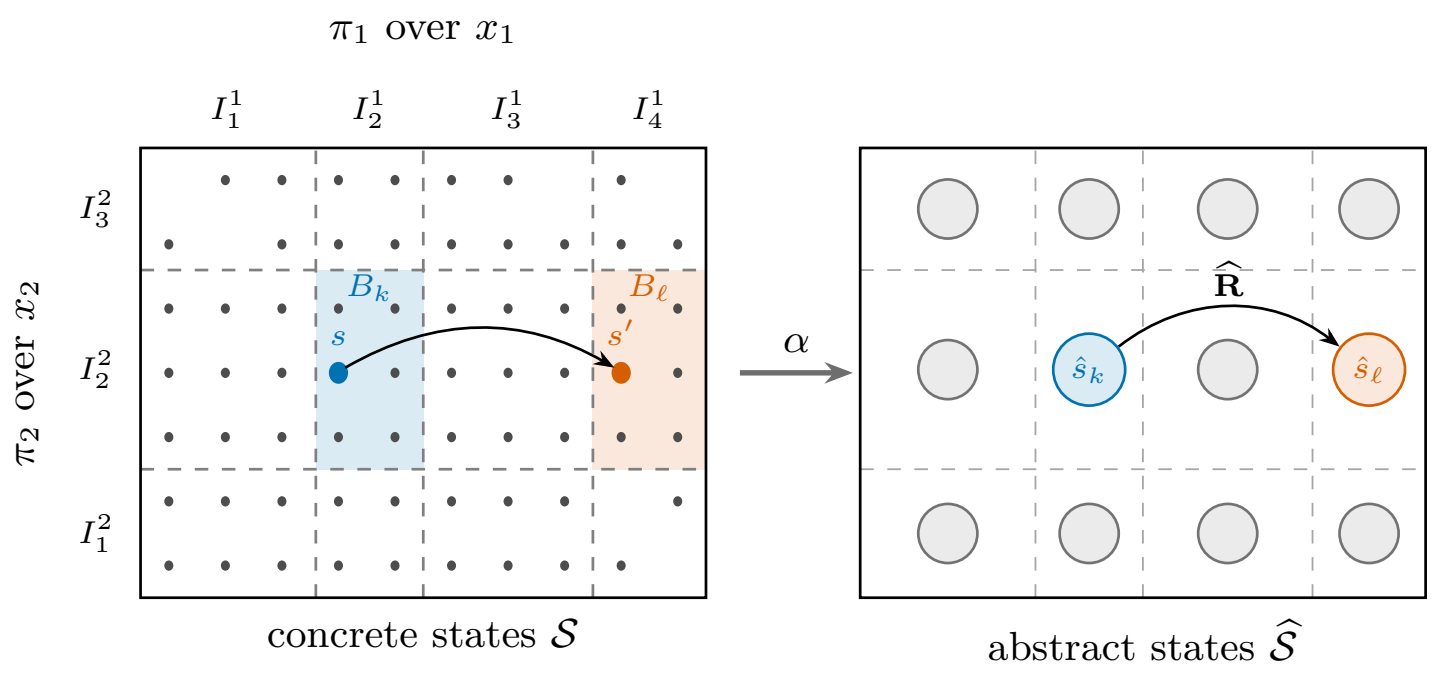

\begin{definition}[Abstraction predicate \leanformalized{https://anonymous.4open.science/r/hm-mean-9D42/lean/HMAbs/Defs/Predicate.lean}]\label{def:predicate}
An \emph{abstraction predicate} $\pi_i$ over variable $x_i$ is a
partition of its integer domain $D_i$ into finitely
many disjoint, contiguous intervals; $I^i_j\subseteq D_i$ is the
$j$-th interval of $\pi_i$. A
\emph{predicate set} $\Pi=\{\pi_1,\ldots,\pi_n\}$ contains one predicate
per variable. A non-abstracted variable is partitioned into
singletons. A \emph{cut point} of $\pi_i$ is a boundary between two of its
intervals (Fig.~\ref{fig:intervals}).
\end{definition}

\begin{definition}[Blocks \leanformalized{https://anonymous.4open.science/r/hm-mean-9D42/lean/HMAbs/Defs/Blocks.lean}]\label{def:blocks}
The \emph{blocks} $\mathcal B=\{B_1,\ldots,B_M\}$ are the nonempty sets
$\Sset\cap(I^1_{j_1}\times\cdots\times I^n_{j_n})$, where $I^i_{j_i}$ is an
interval of $\pi_i$ for each $i=1,\ldots,n$.
The blocks are pairwise disjoint and their
union is $\Sset$. We write $N_k=|B_k|$ for the number of concrete states in
$B_k$.
The number $M$ of blocks counts only \emph{nonempty} boxes, hence
$M\le\prod_{i=1}^n |\pi_i|$ (Fig.~\ref{fig:intervals}).
\end{definition}

\begin{definition}[Abstract states \leanformalized{https://anonymous.4open.science/r/hm-mean-9D42/lean/HMAbs/Defs/AbstractStates.lean}]\label{def:abstract-states}
Each abstract state in the abstract state space $\Shat=\{\hat s_1,\ldots,\hat s_M\}$ represents a unique block of concrete states, with \emph{abstraction map} $\alpha(s)=\hat s_k\ \Longleftrightarrow\ s\in B_k$ (Fig.~\ref{fig:intervals}).
Equivalently, $B_k=\alpha^{-1}(\hat s_k)$.
\end{definition}


The partition must respect state labels ($L(s) = L(s')$ for $s, s' \in B_k$), enforced by placing cut points at proposition thresholds. The abstract initial state $\hat s_0=\alpha(s_0)$ is the block containing $s_0$.
In the abstraction, concrete transitions lead into well-defined blocks without a notion of individual states within blocks:

\begin{definition}[Block-level rate \leanformalized{https://anonymous.4open.science/r/hm-mean-9D42/lean/HMAbs/Defs/BlockRate.lean}]\label{def:block-rate}
For a state $s\in B_k$ and a block $B_\ell$, the \emph{block-level rate}
from $s$ into $B_\ell$ is
$\rkl{k}{\ell}{s}=\sum_{s'\in B_\ell}\R(s,s')$ for $\ell=1,\ldots,M$.
\end{definition}

\begin{figure}[t]
 \centering
\begin{tikzpicture}[hm arrow,scale=0.85,transform shape]
 \node[hm block,minimum width=3.9cm,minimum height=2.7cm,
       label={[font=\footnotesize,anchor=south west,inner sep=2pt]north west:$B_k$}] (Bk) at (0,0) {};
 \node[hm block,minimum width=1.7cm,minimum height=2.3cm,
       label={[font=\footnotesize]right:$B_\ell$}] (Bl) at (6.6,1.5) {};
 \node[hm block,minimum width=1.7cm,minimum height=1.3cm,
       label={[font=\footnotesize]right:$B_m$}] (Bm) at (6.6,-1.5) {};
 \node[hm state] (s)  at (0.9,0)      {$s$};
 \node[hm state] (s1) at (-1.1,0.7)   {$s'$};
 \node[hm state] (s2) at (-1.1,-0.7)  {$s''$};
 \node[hm state] (u1) at (6.6,2.0)    {$u_1$};
 \node[hm state] (u2) at (6.6,1.0)    {$u_2$};
 \node[hm state] (v1) at (6.6,-1.5)   {$v_1$};
 \draw[hm internal] (s) -- node[pos=0.5,above,sloped,font=\scriptsize,inner sep=1.5pt]{$\R(s,s')$} (s1);
 \draw[hm internal] (s) -- node[pos=0.5,below,sloped,font=\scriptsize,inner sep=1.5pt]{$\R(s,s'')$} (s2);
 \draw[hm arrow,hmblue] (s) -- node[pos=0.5,above,sloped,font=\scriptsize,inner sep=1.5pt]{$\R(s,u_1)$} (u1);
 \draw[hm arrow,hmblue] (s) -- node[pos=0.5,below,sloped,font=\scriptsize,inner sep=1.5pt]{$\R(s,u_2)$} (u2);
 \draw[hm arrow,hmorange] (s) -- node[pos=0.5,below,sloped,font=\scriptsize,inner sep=1.5pt]{$\R(s,v_1)$} (v1);
 \node[font=\footnotesize,hmgray,inner sep=1pt,anchor=north] at (0,-1.5)
   {$\rkl{k}{k}{s}=\R(s,s')+\R(s,s'')$};
 \node[font=\footnotesize,hmblue,inner sep=1pt,anchor=south] at (3.7,2.75)
   {$\rkl{k}{\ell}{s}=\R(s,u_1)+\R(s,u_2)$};
 \node[font=\footnotesize,hmorange,inner sep=1pt,anchor=north] at (5.3,-2.6)
   {$\rkl{k}{m}{s}=\R(s,v_1)$};
\end{tikzpicture}
 \caption{Block-level rates of a state $s\in B_k$. Each block-level rate is the
 \emph{sum} of the concrete rates $\R(s,\cdot)$ on the arrows of its group:
 $\rkl{k}{k}{s}$ for the internal transitions (dotted), $\rkl{k}{\ell}{s}$ for the
 transitions into $B_\ell$ (blue), and $\rkl{k}{m}{s}$ for the transition into $B_m$
 (orange). By Lemma~\ref{lem:block-total}, the three sums add up to $E(s)$.}
 \label{fig:blocklevel}
\end{figure}
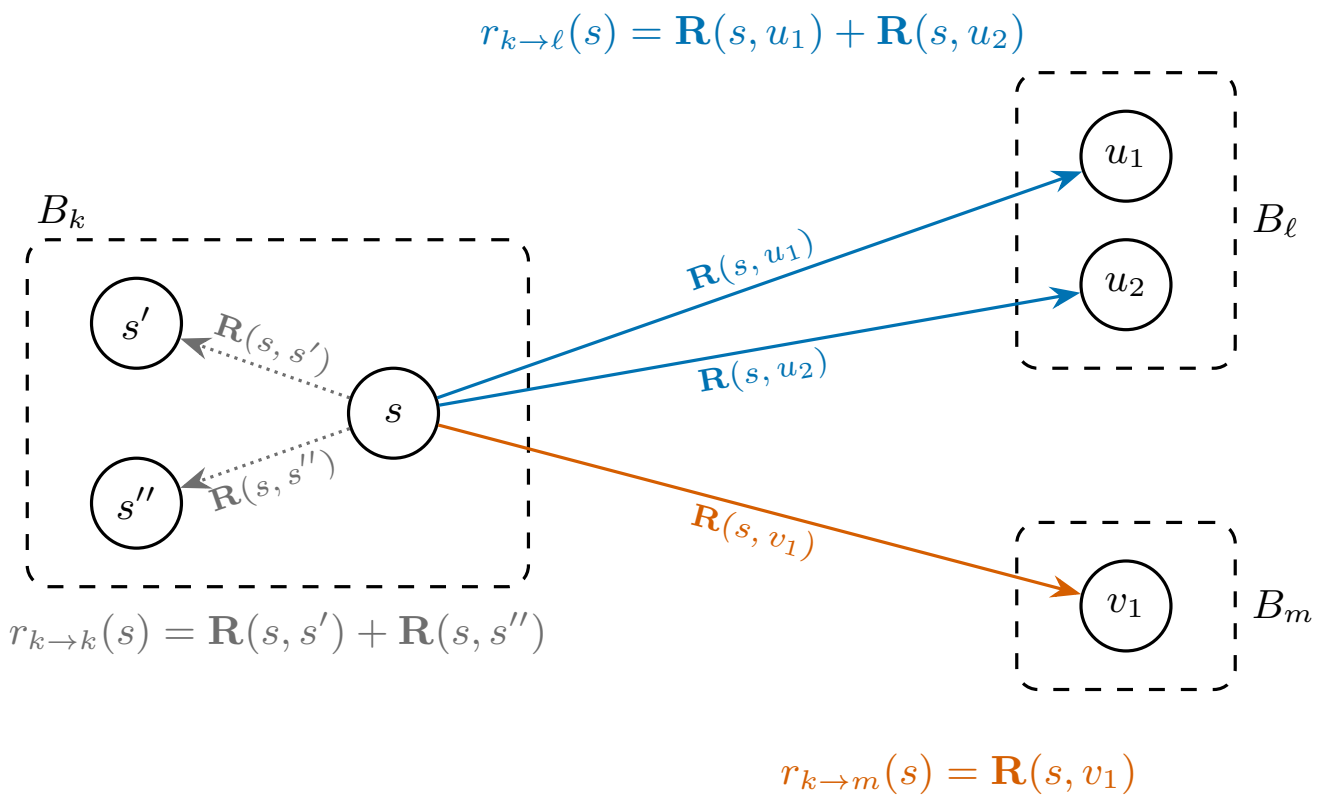

In the case of equivalent source and destination blocks,
$\rkl{k}{k}{s}$ is the total rate of \emph{internal} transitions, i.e.\ of
transitions from $s$ to \emph{other concrete states inside $B_k$}. Viewed
at the concrete level, such a transition is an ordinary transition between two
different states, so $\R(s,s')$ may be positive for $s,s'\in B_k$ and
$\rkl{k}{k}{s}$ need not be zero, even though $\R(s,s)=0$. Viewed at the
abstract level, however, the block $B_k$ is a single abstract state
$\hat s_k$, and an internal transition would be a self-loop of $\hat s_k$,
which a CTMC does not allow; the abstract rate $\Rhat(\hat s_k,\hat s_k)$
must therefore be zero (Sect.~\ref{sec:abstract}).
Figure~\ref{fig:blocklevel} illustrates the three
kinds of outgoing transitions of a state $s\in B_k$: internal ones, ones
into a block $B_\ell$, and ones into another block $B_m$.

\begin{lemma}[Block totals \leanproof{https://anonymous.4open.science/r/hm-mean-9D42/lean/HMAbs/Lemmas/BlockTotal.lean}]\label{lem:block-total}
For every $s\in B_k$, $\sum_{\ell=1}^M \rkl{k}{\ell}{s}=E(s)$, and if $E(s)>0$, then
$\PP(\alpha(\nu_s)=\hat s_\ell)=\rkl{k}{\ell}{s}/E(s)$ for every $\ell$.
\end{lemma}

See Appendix~\ref{app:proof-block-total} for the proof for Lemma~\ref{lem:block-total}. 
The index $\ell$ in Lemma~\ref{lem:block-total} runs over \emph{all} $M$
blocks, including $\ell=k$. The distinction between a concrete
transition and an exit from a block is essential for the
aggregation in Sect.~\ref{sec:hm}.

We use two time notions throughout.
The \emph{residence time} is the duration spent in a state or block until exit; for concrete state $s$, this is $\tau_s$. The \emph{waiting time} is the duration until the next \emph{concrete} transition, a concept limited to concrete states (where it equals $\tau_s$) or blocks under the uniform experiment (Sect.~\ref{sec:derive}), where the next transition may be internal. Abstract states have only residence times. For a block, experimental waiting time and residence time differ because block residence may span multiple internal transitions.

\section{Harmonic-Mean Aggregation}\label{sec:hm}

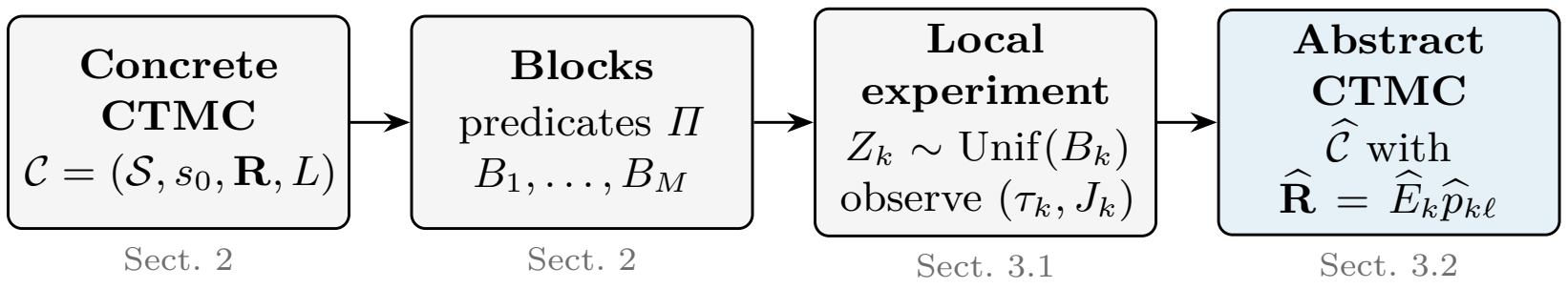
\begin{figure}[t]
 \centering
\scalebox{0.92}{%
\begin{tikzpicture}[node distance=0.45cm]
 \node[hm box] (c) {\textbf{Concrete\\CTMC}\\[1pt]
   $\C=(\Sset,s_0,\R,L)$};
 \node[hm box,right=of c] (b) {\textbf{Blocks}\\[1pt]
   predicates $\Pi$\\ $B_1,\dots,B_M$};
 \node[hm box,right=of b] (l) {\textbf{Local\\experiment}\\[1pt]
   $Z_k\sim\Unif(B_k)$\\ observe $(\tau_k,J_k)$};
 \node[hm box,right=of l,fill=hmblue!10] (a) {\textbf{Abstract\\CTMC}\\[1pt]
   $\Chat$ with\\ $\Rhat=\Ehat_k\phat_{k\ell}$};
 \draw[hm arrow] (c) -- (b);
 \draw[hm arrow] (b) -- (l);
 \draw[hm arrow] (l) -- (a);
 \node[hm label,hmgray,below=2pt of c] {Sect.~\ref{sec:ctmc}};
 \node[hm label,hmgray,below=2pt of b] {Sect.~\ref{sec:predicates}};
 \node[hm label,hmgray,below=2pt of l] {Sect.~\ref{sec:derive}};
 \node[hm label,hmgray,below=2pt of a] {Sect.~\ref{sec:abstract}};
\end{tikzpicture}%
}
 \caption{Overview of the construction. Predicates induce blocks
 (Sect.~\ref{sec:predicates}). Inside each block we study a local
 experiment: a state is drawn uniformly and its next transition is
 observed (Sect.~\ref{sec:derive}). Matching the mean waiting time
 and the destination law of this experiment yields the abstract rates
 (Sect.~\ref{sec:abstract}).}
 \label{fig:pipeline}
\end{figure}

The partition of Sect.~\ref{sec:predicates} fixes which concrete states are
grouped together. It remains to define the transition rates between
the resulting abstract states. Our construction has two steps, matching
(i)~the \emph{mean waiting time until the next concrete transition} and (ii)~the
\emph{probability of its destination block}, in a local experiment
inside each block (Fig.~\ref{fig:pipeline}).
Sect.~\ref{sec:derive} derives the construction,
and Sect.~\ref{sec:abstract} defines the abstract CTMC. A small
running example, introduced in Sect.~\ref{sec:derive}, accompanies every step.
Symbols are listed in Table~\ref{tab:notation} in
Appendix~\ref{app:notation}.

\begin{convention}[ \leanformalized{https://anonymous.4open.science/r/hm-mean-9D42/lean/HMAbs/Defs/Convention.lean}]\label{conv:blocks}
Blocks are partitioned beforehand so that states within each block have either all positive exit rates ($E(s)>0$) or all $E(s)=0$. Absorbing blocks ($E(s)=0$) map to absorbing abstract states with zero outgoing rates; we derive formulas for blocks with positive exit rates.
\end{convention}

\subsection{From waiting times to the harmonic mean}\label{sec:derive}

Consider the following local experiment.

\begin{definition}[Local uniform experiment \leanformalized{https://anonymous.4open.science/r/hm-mean-9D42/lean/HMAbs/Defs/LocalExperiment.lean}]\label{def:local}
Let $B_k$ be a block with positive exit rates. Draw a state $Z_k$ uniformly from $B_k$, i.e.\ $\PP(Z_k=s)=1/N_k$ for $s\in B_k$,
and observe its next concrete transition. Let $\tau_k$ be the
waiting time and $J_k\in\{1,\ldots,M\}$ the index of the destination
block. Thus, conditionally on $Z_k=s$, the waiting time has the law of
$\tau_s$, and $J_k=\ell$ exactly when $\nu_s\in B_\ell$. The distribution
$\ell\mapsto\PP(J_k=\ell)$ of $J_k$ is the \emph{destination law} of $B_k$.
\end{definition}

The uniform law of Def.~\ref{def:local} is a \emph{reference
distribution} used to define the aggregation. It does not assert that
the original CTMC enters $B_k$ uniformly, and it does not make the
initial state random: the concrete chain $\C$ still starts in the single state $s_0$.
Its abstract counterpart $\hat s_0=\alpha(s_0)$ is the block that contains $s_0$, so
in $\Chat$ the state $s_0$ is merged with the other states of that block, as for any abstraction.
The error bound of Sect.~\ref{sec:error} accounts for this through the weight of $s_0$ within
its block.
We illustrate this procedure on a small running example. 

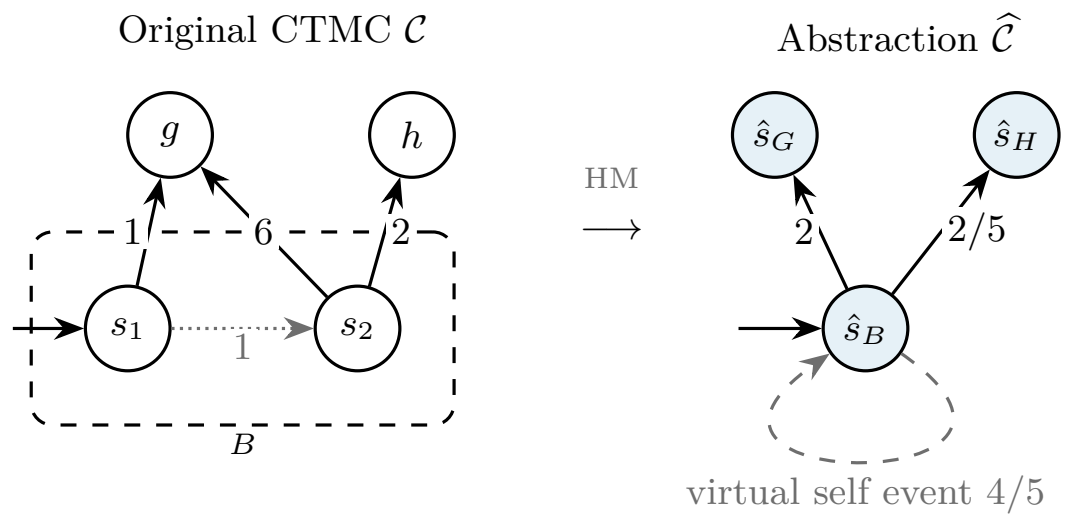
\begin{figure}[tb]
 \centering
\begin{tikzpicture}[hm arrow,scale=0.8,transform shape]
 \node[hm state] (s1) at (0,0)      {$s_1$};
 \node[hm state] (s2) at (1.9,0)    {$s_2$};
 \node[hm state] (g)  at (0.35,1.6) {$g$};
 \node[hm state] (h)  at (2.35,1.6) {$h$};
 \draw[hm internal] (s1) -- node[below,hm rate] {$1$} (s2);
 \draw[hm arrow] (s1) -- node[midway,left,hm rate]  {$1$} (g);
 \draw[hm arrow] (s2) -- node[midway,hm rate] {$6$} (g);
 \draw[hm arrow] (s2) -- node[midway,right,hm rate] {$2$} (h);
 \begin{scope}[on background layer]
  \node[hm block,fit=(s1)(s2),inner sep=3.5mm,label={[hm label]below:$B$}] {};
 \end{scope}
 \draw[hm arrow] (-0.95,0) -- (s1);
 \node[font=\small] at (1.2,2.45) {Original CTMC $\C$};
 \node at (4.0,0.8) {$\longrightarrow$};
 \node[hm label,hmgray,align=center] at (4.0,1.25) {HM};
 \node[hm abs state] (ab) at (6.1,0)    {$\hat s_B$};
 \node[hm abs state] (ag) at (5.35,1.6) {$\hat s_G$};
 \node[hm abs state] (ah) at (7.35,1.6) {$\hat s_H$};
 \draw[hm arrow] (ab) -- node[midway,left,hm rate]  {$2$} (ag);
 \draw[hm arrow] (ab) -- node[midway,right,hm rate] {$2/5$} (ah);
 \draw[hm arrow] (5.05,0) -- (ab);
 \draw[hm arrow,dashed,hmgray] (ab) to[out=-35,in=-145,looseness=9]
   node[below,font=\footnotesize,hmgray] {virtual self event $4/5$} (ab);
 \node[font=\small] at (6.35,2.45) {Abstraction $\Chat$};
\end{tikzpicture}
 \caption{The original and the abstract CTMC of the running example. Edge labels are rates.
 The internal transition $s_1\to s_2$ (dotted) gives no transition of $\hat s_B$; it only
 contributes the virtual self event of rate $\Ehat_B\phat_{BB}=4/5$ (dashed loop, not part of $\Chat$).}
 \label{fig:example}
\end{figure}

\begin{exmp}
Let $\Sset=\{s_1,s_2,g,h\}$ with
initial state $s_0=s_1$ and nonzero rates
$\R(s_1,s_2)=1$, $\R(s_1,g)=1$, $\R(s_2,g)=6$, and $\R(s_2,h)=2$.
States $g$ and $h$ are absorbing. With $\AP=\{a,b\}$, let
$L(s_1)=L(s_2)=\varnothing$, $L(g)=\{a\}$, and $L(h)=\{b\}$. Encode the
states by one variable $x\in\{0,1,2,3\}$ with $s_1=0$, $s_2=1$, $g=2$,
$h=3$, and take the predicate $\pi=\{[0,1],[2,2],[3,3]\}$. It yields three
blocks, $B=\{s_1,s_2\}$, $G=\{g\}$, and $H=\{h\}$, with abstract states
$\hat s_B$, $\hat s_G$, $\hat s_H$ and $\hat s_0=\hat s_B$. Here
we write block names instead of numerical indices as subscripts. Only $B$ has
positive exit rates, $E(s_1)=2$ and $E(s_2)=8$, and the transition
$s_1\to s_2$ is internal to $B$. Figure~\ref{fig:example} shows the
concrete chain $\C$ and, as a preview, its abstraction $\Chat$; the rates of
$\Chat$ are derived in the steps below and in Sect.~\ref{sec:abstract}.
\end{exmp}

\paragraph{Step 1: average the waiting times.}
In concrete state $s$, the mean residence time is $1/E(s)$. Averaging under Def.~\ref{def:local} yields the block's mean waiting time:
$\EE[\tau_k] = \sum_{s\in B_k}\PP(Z_k=s)\,\EE[\tau_k\mid Z_k=s] = \frac{1}{N_k}\sum_{s\in B_k}\frac{1}{E(s)}$.
Thus, waiting times (rather than rates) are averaged.

\begin{exmp}
For $B$,
$\EE[\tau_B]=\frac12\bigl(\frac12+\frac18\bigr)=\frac5{16}$.
\end{exmp}

\paragraph{Step 2: choose one exponential clock with that mean.}
Let $\hat\tau_k$ be an auxiliary abstract event clock, exponentially
distributed with a rate $\Ehat_k>0$, i.e.\ $\hat\tau_k\sim\Exp(\Ehat_k)$.
We call $\Ehat_k$ the abstract \emph{event rate}; it is still to be chosen.
The mean of $\hat\tau_k$ is $1/\Ehat_k$, so matching the mean $\EE[\tau_k]$ of Step~1
requires $1/\Ehat_k=\frac1{N_k}\sum_{s\in B_k}1/E(s)$, hence
$\Ehat_k=\bigl(\frac1{N_k}\sum_{s\in B_k}1/E(s)\bigr)^{-1}$.
Thus, the event rate is the HM of the total concrete exit
rates. It counts events whose
destination may lie inside $B_k$, so it is not necessarily the rate of
leaving the abstract state $\hat s_k$ (see
Sect.~\ref{sec:abstract}).

\begin{exmp}
For $B$, $\Ehat_B=(5/16)^{-1}=16/5$.
\end{exmp}

Averaging waiting times yields the harmonic mean rate $\Ehat_k$, whereas the \emph{arithmetic mean} (AM) rate $\Eam_k = N_k^{-1}\sum_{s\in B_k} E(s)$ yields a strictly smaller mean residence time ($1/\Eam_k \le 1/\Ehat_k$ by the AM--HM inequality, with equality iff rates in $B_k$ coincide). An AM rate is thus faster than the concrete chain. HM is the unique choice matching the mean waiting time (Prop.~\ref{prop:preserve}) and providing the generator weights $w_k$ used for error analysis (Prop.~\ref{prop:structure}, Sect.~\ref{sec:error}).


\paragraph{Step 3: average the next-event destinations.}
From state $s \in B_k$, the probability of entering $B_\ell$ is $\rkl{k}{\ell}{s}/E(s) = \sum_{s'\in B_\ell}\R(s,s')/E(s)$ (Lemma~\ref{lem:block-total}). Averaging under the uniform choice yields
$\phat_{k\ell} = \frac1{N_k}\sum_{s\in B_k}\rkl{k}{\ell}{s}/E(s)$,
the probability that the local experiment's next concrete transition enters $B_\ell$. This includes internal transitions ($\ell=k$). The abstract destination index $\hat J_k$ is drawn independently of $\hat\tau_k$ with $\PP(\hat J_k=\ell)=\phat_{k\ell}$, acting as the abstract counterpart of $J_k$ (Def.~\ref{def:local}).

\begin{exmp}
Block-level rates and the next-event probabilities in $B$ are as follows:

{\centering
\setlength{\tabcolsep}{4pt}
\renewcommand{\arraystretch}{1.05}
\begin{tabular}{c ccc c ccc}
\toprule
 & \multicolumn{3}{c}{Block-level rates} & &
 \multicolumn{3}{c}{Next-event probabilities}\\
\cmidrule(r){2-4}\cmidrule(l){6-8}
State & $B$ & $G$ & $H$ & $E(s)$ & $B$ & $G$ & $H$\\
\midrule
$s_1$ & $1$ & $1$ & $0$ & $2$ & $1/2$ & $1/2$ & $0$\\
$s_2$ & $0$ & $6$ & $2$ & $8$ & $0$ & $3/4$ & $1/4$\\
\bottomrule
\end{tabular}\par}
\smallskip
Averaging each probability column gives
$\phat_{BB}=\frac14$, $\phat_{BG}=\frac58$, and $\phat_{BH}=\frac18$.
\end{exmp}

\subsection{The abstract CTMC}\label{sec:abstract}

\begin{definition}[Harmonic-mean abstraction \leanformalized{https://anonymous.4open.science/r/hm-mean-9D42/lean/HMAbs/Defs/HMAbstraction.lean}]\label{def:hm}
The \emph{harmonic-mean (HM) abstraction} of $\C$ with respect to $\Pi$
is the CTMC $\Chat=(\Shat,\hat s_0,\Rhat,\Lhat)$ with states, initial state, and labels as
in Def.~\ref{def:abstract-states} and Sect.~\ref{sec:predicates} and, for a
positive-rate source block $B_k$, the rates are
$\Rhat(\hat s_k,\hat s_\ell)=\Ehat_k\phat_{k\ell}$ for $\ell\ne k$ and
$\Rhat(\hat s_k,\hat s_k)=0$.
For a block consisting entirely of absorbing states, the corresponding
row of $\Rhat$ is zero; the
divisions in Steps~2 and~3 are not used for it.
\end{definition}

Definition~\ref{def:hm} assigns to $B_k$ a \emph{single} event rate,
the harmonic mean $\Ehat_k$ of the total exit rates $E(s)$ (Step~2), and splits it
among the destination blocks with the averaged branching probabilities
$\phat_{k\ell}$ (Step~3), which gives
$\Rhat(\hat s_k,\hat s_\ell)=\Ehat_k\phat_{k\ell}$.
It does \emph{not} take a separate harmonic mean of
$\rkl{k}{\ell}{s}$ over $s\in B_k$ for each destination block $B_\ell$.
The two constructions differ, because the harmonic mean does not commute with
addition: the harmonic mean of the sums
$E(s)=\sum_{\ell}\rkl{k}{\ell}{s}$ is in general not the sum of the harmonic
means of the summands.
In the running example, the separate harmonic means over
$s_1,s_2$ would be $2/(1/1+1/6)=12/7$ for $G$, and $0$ for $H$
(with the convention that the harmonic mean is $0$ if some rate is zero,
here $\rkl{B}{H}{s_1}=0$). Counting $B$ itself, which gets $0$ for the same reason, they add up to
$12/7<16/5=\Ehat_B$,
so the mean waiting time of Step~2 would no longer be matched.

\paragraph{Event rate versus abstract exit rate.}
The total rate of leaving an abstract state is
$\Eout_k\defeq\sum_{\ell\ne k}\Rhat(\hat s_k,\hat s_\ell)=\Ehat_k(1-\phat_{kk})$
for a positive-rate block.

The remaining part $\Ehat_k\phat_{kk}$ of the event rate corresponds to
events that do not change the abstract state; we call such events
\emph{virtual self events}. They stay in the branching distribution but
do not appear in $\Rhat$, so $\Ehat_k=\Eout_k+\Ehat_k\phat_{kk}$.
The abstract generator is therefore $\Qhat_{k\ell}=\Rhat(\hat s_k,\hat s_\ell)$ for
$\ell\ne k$ and $\Qhat_{kk}=-\Eout_k$.
For an absorbing block, $\Eout_k=0$ and the generator row is zero.

\begin{exmp}
The abstract rates out of $\hat s_B$ are
$\Rhat(\hat s_B,\hat s_G)=\frac{16}{5}\cdot\frac58=2$ and
$\Rhat(\hat s_B,\hat s_H)=\frac{16}{5}\cdot\frac18=\frac25$, so
$\Eout_B=2+\frac25=\frac{12}5$, not $\Ehat_B=\frac{16}5$. The remaining part
$\Ehat_B\phat_{BB}=\frac45$ of the event rate is the virtual self event, and
$\Ehat_B=\frac{16}5=2+\frac25+\frac45$. The abstract labels are inherited from
the blocks, $\Lhat(\hat s_B)=\varnothing$, $\Lhat(\hat s_G)=\{a\}$ and
$\Lhat(\hat s_H)=\{b\}$. These are the values drawn in Fig.~\ref{fig:example}. In
the state orders $(s_1,s_2,g,h)$ and $(\hat s_B,\hat s_G,\hat s_H)$, the
generators have the rows
$Q_{s_1\cdot}=(-2,1,1,0)$, $Q_{s_2\cdot}=(0,-8,6,2)$, and zero rows for $g$ and $h$,
and $\Qhat_{B\cdot}=(-12/5,2,2/5)$ and zero rows for $\hat s_G$ and $\hat s_H$.
The mean time to the next event is $\EE[\hat\tau_B]=5/16$, whereas the mean
residence time in $\hat s_B$ is $\EE[\hat\tau_B^{\mathrm{out}}]=5/12$.
\end{exmp}

\begin{lemma}[Well-definedness \leanproof{https://anonymous.4open.science/r/hm-mean-9D42/lean/HMAbs/Lemmas/WellDefinedness.lean}]\label{lem:wd}
Under Convention~\ref{conv:blocks}, let $\Qhat$ be the matrix
of Sect.~\ref{sec:abstract}.
\begin{enumerate}
\item[(a)] $\Qhat$ is a generator. Hence $\Chat$ is a finite CTMC.
\item[(b)] If $\Eout_k>0$, then in $\Chat$ the residence time $\tauout$ in
 $\hat s_k$ and the next abstract state are independent, with
 $\tauout\sim\Exp(\Eout_k)$ and
 $\PP(\text{next abstract state}=\hat s_\ell\mid\hat s_k)
 =\phat_{k\ell}/(1-\phat_{kk})$ for $\ell\ne k$.
\item[(c)] Let $B_k$ have positive exit rates, so that $\phat_{kk}$ is
 defined. If $\phat_{kk}=1$, then $\hat s_k$ is absorbing in $\Chat$, even
 though concrete transitions may continue inside $B_k$. If $\phat_{kk}=0$,
 then $\Ehat_k=\Eout_k$.
\end{enumerate}
\end{lemma}

Lemma~\ref{lem:wd} is proven in Appendix~\ref{app:proof-wd}.
Statement~\ref{lem:wd}(a) concerns the local uniform experiment of
Definition~\ref{def:local}. Its abstract counterpart is the auxiliary
event clock $\hat\tau_k\sim\Exp(\Ehat_k)$ of Step~2 together with the
destination index $\hat J_k$ of Step~3.

\begin{proposition}[Local preservation \leanproof{https://anonymous.4open.science/r/hm-mean-9D42/lean/HMAbs/Propositions/LocalPreservation.lean}]\label{prop:preserve}
For a block $B_k$ with $E(s)>0$ for every $s\in B_k$,
HM aggregation preserves the mean waiting time,
$\EE[\hat\tau_k]=\EE[\tau_k]$ (\emph{mean-time preservation}), and the
destination law, $\PP(\hat J_k=\ell)=\PP(J_k=\ell)$ for $\ell=1,\ldots,M$
(\emph{destination-law preservation}).
Moreover, among all exponential clocks $\Exp(\lambda)$ with $\lambda>0$, only
$\lambda=\Ehat_k$ satisfies mean-time preservation. In this sense, the harmonic
mean is forced by matching the mean waiting time.
\end{proposition}

Proposition~\ref{prop:preserve} is proven in Appendix~\ref{app:proof-preserve}.
Next we express $\Rhat$ directly through the concrete rates
$\rkl{k}{\ell}{s}$, using the following weights.

\begin{definition}[Sojourn-time weights \leanformalized{https://anonymous.4open.science/r/hm-mean-9D42/lean/HMAbs/Defs/Weights.lean}]\label{def:weights}
Let $B_k$ be a block with positive exit rates. The \emph{sojourn-time weight} of
a state $s\in B_k$ is
$w_k(s)=E(s)^{-1}\big/\sum_{s'\in B_k}E(s')^{-1}$,
so that $w_k(s)>0$ and $\sum_{s\in B_k}w_k(s)=1$.
\end{definition}

By Step~2, equivalently, $w_k(s)=\frac1{N_k}\cdot\frac{\Ehat_k}{E(s)}$, the
uniform weight $1/N_k$ rescaled by $\Ehat_k/E(s)$. A state whose exit rate equals
the harmonic mean $\Ehat_k$ carries exactly the uniform weight; slower states
carry more, faster states less.

\begin{proposition}[Structure of the abstract rates \leanproof{https://anonymous.4open.science/r/hm-mean-9D42/lean/HMAbs/Propositions/Structure.lean}]\label{prop:structure}
Let $B_k$ be a block with positive exit rates.
\begin{enumerate}
\item[(a)] \emph{(Time-weighted average.)}
 The weights $w_k$ of Def.~\ref{def:weights} form a probability
 distribution on $B_k$, and
 $\Rhat(\hat s_k,\hat s_\ell)=\sum_{s\in B_k}w_k(s)\,\rkl{k}{\ell}{s}$ for $\ell\ne k$,
 $\Ehat_k=\sum_{s\in B_k}w_k(s)\,E(s)$,
 so $\Eout_k=\sum_{s\in B_k}w_k(s)\sum_{\ell\ne k}\rkl{k}{\ell}{s}$.
 More generally, $\Ehat_k\phat_{k\ell}=\sum_{s\in B_k}w_k(s)\,\rkl{k}{\ell}{s}$ for
 all $\ell$, including $\ell=k$.
\item[(b)] \emph{(Singletons.)} If $N_k=1$, then
 $\Rhat(\hat s_k,\hat s_\ell)=\rkl{k}{\ell}{s}$ for $\ell\ne k$, where
 $B_k=\{s\}$. If all blocks are singletons, then $\alpha$ is a bijection
 $\Sset\to\Shat$ with $\alpha(s_0)=\hat s_0$, $\Lhat(\alpha(s))=L(s)$ and
 $\Rhat(\alpha(s),\alpha(s'))=\R(s,s')$ for all $s,s'\in\Sset$. Thus $\Chat$ is
 $\C$ with its states renamed by $\alpha$.
\item[(c)] \emph{(Lumpability.)} If for every $\ell\ne k$ the rate
 $\rkl{k}{\ell}{s}=c_{k\ell}$ is the same for all $s\in B_k$, then
 $\Rhat(\hat s_k,\hat s_\ell)=c_{k\ell}$ for all $\ell\ne k$.
\end{enumerate}
\end{proposition}

Proposition~\ref{prop:structure} is proven in Appendix~\ref{app:proof-structure}.
While Proposition~\ref{prop:preserve} describes the abstraction through the local
experiment, part~(a) uses rates directly: $\Rhat$ is a weighted average of the
concrete rates $\rkl{k}{\ell}{s}$, in which a state is weighted by its mean
residence time $1/E(s)$: states in which the chain lingers contribute more. The hypothesis of part~(c) is
the standard condition of ordinary lumpability~\cite{Buchholz1994,CardelliEtAl2023}:
for every block $B_\ell\ne B_k$, all states of $B_k$ have the same total rate into
$B_\ell$. It requires
no assumption on the rate $\rkl{k}{k}{s}$ of internal transitions
or on the exit rates $E(s)$, which may differ between states.

\begin{exmp}
As above, $w_B=(4/5,1/5)$, and
$\Rhat(\hat s_B,\hat s_G)=\tfrac45\cdot1+\tfrac15\cdot6=2$ by~\ref{prop:structure}(a).
\end{exmp}

Proposition~\ref{prop:preserve} preserves a mean waiting time and a
destination law, not the whole stochastic behavior of the block. The
concrete waiting-time survival function is a mixture of exponentials,
$\PP(\tau_k>t)=\frac1{N_k}\sum_{s\in B_k}e^{-E(s)t}$, which is generally not
the single exponential $\PP(\hat\tau_k>t)=e^{-\Ehat_k t}$ ($t\ge0$).
Figure~\ref{fig:survival} plots both laws for the block of the running example:
the concrete mixture (solid) and the HM clock (dashed) enclose the same area
(the same mean) but have different shapes. The dotted curve, the clock with the
arithmetic-mean rate, is shown for contrast; its mean is smaller. The joint distribution of $(\tau_k,J_k)$ is not
preserved either. In the local experiment, both $\tau_k$ and $J_k$ depend on the
drawn state $s$, so they are in general dependent: for example, a state with a
large $E(s)$ has short waiting times and may favor a particular destination block.
In $\Chat$, by construction, $\hat\tau_k$ and $\hat J_k$ are independent.
The dependence arises only because the experiment mixes states with different
rates; within a single concrete state $s$, $\tau_s$ and $\nu_s$ are independent.
A CTMC cannot represent such a dependence between the next-event time and its
destination, so $\Chat$ cannot reproduce it, which is one reason why $\Chat$ is
only an approximation (Sect.~\ref{sec:error}).

Most importantly, the waiting time until the next concrete transition
is not the residence time in $B_k$. An internal transition changes the concrete
state and thus the subsequent rates, and $\Chat$ does not record the
new concrete state. Proposition~\ref{prop:preserve} therefore does
not imply preservation of block residence times or of time-bounded
reachability probabilities. It also neither assumes nor establishes
uniform entry into $B_k$ in the original CTMC. In the special
cases of Proposition~\ref{prop:structure}(b,c), the rates of
$\Chat$ are exactly those of $\C$ (singleton blocks) or of its lumped
chain.

Let $T_B$ be the residence time in $B$ in $\C$, i.e.\ the time until $\C$
first leaves $B$. From $s_2$, its mean
is $1/8$. From $s_1$, it is $1/2+\tfrac12\cdot\tfrac18=9/16$: after
the first wait, the chain moves to $s_2$ with probability $1/2$ and
waits again. Under the uniform start used in the local experiment, the
mean is $\EE[T_B]=\tfrac12\bigl(\tfrac9{16}+\tfrac18\bigr)=\tfrac{11}{32}$.
Table~\ref{tab:example-compare} compares several quantities of $\C$ and
$\Chat$.

\begin{figure}[t]
\begin{minipage}[c]{0.38\textwidth}
 \centering
\begin{tikzpicture}
 \begin{axis}[
   width=0.80\linewidth,height=3.4cm,
   xmin=0,xmax=1.2,ymin=0,ymax=1.02,
   xlabel={$t$},ylabel={survival prob.},
   xlabel style={font=\footnotesize},ylabel style={font=\footnotesize},
   tick label style={font=\scriptsize},
   grid=major,grid style={gray!25},]
  \addplot[hmblue,very thick,domain=0:1.2,samples=400]
    {0.5*exp(-2*x)+0.5*exp(-8*x)};
  \addplot[hmorange,very thick,dashed,domain=0:1.2,samples=400]
    {exp(-3.2*x)};
  \addplot[hmgreen,very thick,dotted,domain=0:1.2,samples=400]
    {exp(-5*x)};
 \end{axis}
\end{tikzpicture}
\end{minipage}\hfill
\begin{minipage}[c]{0.60\textwidth}
 \centering\scriptsize
 \setlength{\tabcolsep}{3pt}
 \renewcommand{\arraystretch}{1.2}
 \begin{tabular}{@{}>{\raggedright\arraybackslash}p{3.45cm}ccc@{}}
 \toprule
 Quantity (unif.\ start) & $\C$ & $\Chat$ & Match\\
 \midrule
 mean time to next event & $5/16$ & $5/16$ & yes\\
 next-block probs.\ $(B,G,H)$ &
 $(\frac14,\frac58,\frac18)$ & $(\frac14,\frac58,\frac18)$ & yes\\
 \midrule
 mean residence time in $B$ & $11/32$ & $5/12$ & no\\
 same, from $s_0=s_1$ & $9/16$ & $5/12$ & no\\
 $\PP(\text{reach } G)$, $T\to\infty$ & $7/8$ & $5/6$ & no\\
 $\PP(\text{reach } H)$, $T\to\infty$ & $1/8$ & $1/6$ & no\\
 \bottomrule
 \end{tabular}
\end{minipage}
\vspace{4pt}

\begin{minipage}[t]{0.38\textwidth}
 \caption{Survival functions of the waiting time in $B$: concrete mixture
 (solid, mean $5/16$), HM clock (dashed, mean $5/16$), AM clock (dotted,
 mean $1/5$).}
 \label{fig:survival}
\end{minipage}\hfill
\begin{minipage}[t]{0.60\textwidth}
 \makeatletter\def\@captype{table}\makeatother
 \caption{Quantities of the example. The first two rows are matched by
 construction; the rest depend on how long the chain stays in $B$. The last
 two rows are limits as $T\to\infty$ from $s_0$ (finite $T$: Sect.~\ref{sec:error}).}
 \label{tab:example-compare}
\end{minipage}
\end{figure}

The HM construction matches the local mean time $5/16$ to the next
concrete transition and the three next-event probabilities. This does
not make the mean residence time in the block exact, even when the local
uniform starting distribution holds, and it does not make
the reachability probabilities exact: as $T\to\infty$, reaching $G$ has
probability $7/8$ from $s_1$, but $2\big/\tfrac{12}5=5/6$ from $\hat s_B$
(Table~\ref{tab:example-compare}). Such discrepancies are inherent in
any abstraction that forgets the concrete state inside a block. The
specified original initial state of the example remains $s_0=s_1$.

\section{A Value-Aware Error Bound}\label{sec:error}

HM aggregation keeps the mean waiting time and the destination
law of each block (Prop.~\ref{prop:preserve}), but it does not keep
reachability probabilities. In the
running example, the probability of eventually reaching $g$ is
$7/8$ in $\C$ and $5/6$ in $\Chat$ (Table~\ref{tab:example-compare}), and the same
mismatch appears at every finite time horizon $T$. The natural question is how far
apart the two can be. We want an answer that uses only $\Chat$ and
numbers we can read off the blocks of $\C$, so that no reachability probability ever
has to be computed on the large state space $\Sset$.

The idea is to treat the error as a dynamical quantity. Both chains evolve by
linear differential equations, so their difference does too. We split that
difference into a part that the abstraction sees and a part that it cannot see, and we
bound each part by a small system with one entry per block. The section follows
this idea. Sect.~\ref{sec:err-setting} fixes the setting and the identity that makes the
HM weights the right choice.
Sect.~\ref{sec:err-equations} derives the exact error equations.
Sect.~\ref{sec:err-bound} replaces the quantities we cannot afford to compute by
block-level quantities, states the bound, and solves it on the running example.
Sect.~\ref{sec:err-eval} explains how to evaluate it in general.
All proofs are in Appendix~\ref{app:proofs}.

\subsection{Setting and the HM identity}\label{sec:err-setting}

We first say what is being compared. Fix a set $\Goal\subseteq\Sset$ of goal states.
As usual, for time-bounded reachability, goal states are absorbing, $E(s)=0$ for
$s\in\Goal$. We also assume that $\Goal$ is a union of blocks, which is automatic when
the goal is given by a label (Sect.~\ref{sec:predicates}). Let $\Goalhat=\{\hat s_k : B_k\subseteq\Goal\}$. The two
vectors of time-bounded reachability probabilities are
$P_t=e^{tQ}\ones_{\Goal}\in[0,1]^{\Sset}$ and
$\Phat_t=e^{t\Qhat}\ones_{\Goalhat}\in[0,1]^{\Shat}$.
Here $\ones_{\Goal}$ is the indicator vector of $\Goal$, i.e.\ $\ones_{\Goal}(s)=1$
if $s\in\Goal$ and $0$ otherwise, and $\ones_{\Goalhat}$ is defined in the same way on $\Shat$.
So $P_t(s)$ is the probability that $\C$, started in $s$, reaches $\Goal$
within time $t$, and $\Phat_t(k)=\Phat_t(\hat s_k)$ is defined similarly for $\Chat$: it is the
probability that $\Chat$, started in $\hat s_k$, reaches the abstract goal $\Goalhat$ within time $t$.
Note that these vectors are
indexed by the \emph{starting} state. They therefore satisfy the backward equations
$P_t'=QP_t$ and $\Phat_t'=\Qhat\Phat_t$, where $Q$ and $\Qhat$ act on the column
vectors $P_t$ and $\Phat_t$, respectively. These are not the forward equations for state
distributions. \emph{Our goal is to bound $|P_T(s_0)-\Phat_T(\hat s_0)|$}.
\begin{exmp}
Take $\Goal=\{g\}$, so $\Goalhat=\{\hat s_G\}$.
The rows of $Q$ give $P_t(s_2)'=-8P_t(s_2)+6$ and $P_t(s_1)'=-2P_t(s_1)+P_t(s_2)+1$, hence
$P_t(s_2)=\tfrac34(1-e^{-8t})$ and $P_t(s_1)=\tfrac78-e^{-2t}+\tfrac18e^{-8t}$. In $\Chat$,
$\Phat_t(B)'=-\tfrac{12}5\Phat_t(B)+2$, so $\Phat_t(B)=\tfrac56(1-e^{-12t/5})$. Here $P_T(s_1)$ and $\Phat_T(B)$ are the probabilities of reaching the goal $g$
(resp.\ $\hat s_G$) within time $T$ from $s_1$ in $\C$ and from $B$ in $\Chat$.
After one time unit ($T=1$) they are $0.740$ and $0.758$, and as $T\to\infty$
they tend to $7/8$ and $5/6$.
\end{exmp}

To compare a vector on $\Sset$ with a vector on $\Shat$ we need two
linear maps between the two spaces.
Being linear, both are identified throughout with their matrices in the
canonical bases: $\Up_{s\ell}=1$ exactly when $\alpha(s)=\ell$, and
$\Av_{ks}=w_k(s)$ for $s\in\Bk$ and $0$ otherwise. The \emph{lifting}
$\Up\colon\mathbb R^{\Shat}\to\mathbb R^{\Sset}$ copies the value of a block to all
its states, $(\Up u)(s)=u(\alpha(s))$\ for every vector $u$ on the blocks
and every state $s\in\Sset$. The \emph{averaging}
$\Av\colon\mathbb R^{\Sset}\to\mathbb R^{\Shat}$ goes the other way and takes the weighted mean over each block:
for a vector $f\in\mathbb R^{\Sset}$ and a block index $k$,
$(\Av f)_k=\sum_{s\in B_k}w_k(s)\,f(s)$,
with the HM weights $w_k(s)\propto E(s)^{-1}$ of Def.~\ref{def:weights},
and with the uniform weight on an absorbing block. The composition $\Av\Up$
is the identity on $\mathbb R^{\Shat}$. We also view the generator $Q$ as the linear map
$f\mapsto Qf$ on $\mathbb R^{\Sset}$. Then $\Av Q\Up$ is the composition that lifts a vector
from $\Shat$ to $\Sset$, applies $Q$, and averages back to $\Shat$, so it is an $M\times M$ matrix.

\begin{exmp}
The exit rates are $E(s_1)=2$ and $E(s_2)=8$. The mean waits $\tfrac12$ and
$\tfrac18$ sum to $\tfrac58$, so $w_B=(\tfrac{1/2}{5/8},\tfrac{1/8}{5/8})=(\tfrac45,\tfrac15)$ and
$(\Av f)_B=\tfrac45f(s_1)+\tfrac15f(s_2)$. The lifting $\Up$ copies the value of $B$ to both $s_1$ and $s_2$.
\end{exmp}

\begin{lemma}[HM identity \leanproof{https://anonymous.4open.science/r/hm-mean-9D42/lean/HMAbs/Lemmas/HmIdentity.lean}]\label{lem:hm-identity}
The matrix product $\Av\,Q\,\Up$, of size $M\times M$, satisfies $\Av\,Q\,\Up=\Qhat$.
\end{lemma}

Lemma~\ref{lem:hm-identity} (proven in Appendix~\ref{app:proof-hm-identity}) shows that the abstract generator can be formed by the average of the concrete generator with HM weights. This is Prop.~\ref{prop:structure}(a)
in matrix form, and it is the reason why we average with the weights $w_k$ and not with
the uniform measure on the block: with uniform weights the identity fails.

\begin{exmp}
For states ordered $(s_1,s_2,g,h)$ and blocks $(B,G,H)$. The same weights applied to the rows $(-2,1,1,0)$ and $(0,-8,6,2)$ of $Q$ give the
row of $B$ in $\Av Q$,
$\bigl(\tfrac45(-2)+\tfrac15\cdot0,\ \tfrac45\cdot1+\tfrac15(-8),\ \tfrac45\cdot1+\tfrac15\cdot6,\
\tfrac15\cdot2\bigr)=\bigl(-\tfrac85,\,-\tfrac45,\,2,\,\tfrac25\bigr)$.
Collapsed onto the blocks $(B,G,H)$, this is $(-\tfrac85-\tfrac45,\,2,\,\tfrac25)=(-\tfrac{12}5,2,\tfrac25)$,
which is $\Qhat_{B\cdot}$. Here and below, the dot stands for the
column index, so $\Qhat_{B\cdot}$ is the row of $B$ in $\Qhat$, and likewise $Q(s,\cdot)$ is the row of $s$.
\end{exmp}

\subsection{Error equations}\label{sec:err-equations}

Where does the error come from? Consider one block $B_k$ over a very short time. A concrete
state $s\in B_k$ sends probability into a block $B_\ell$ at total rate $Q(s,B_\ell)=\sum_{s'\in B_\ell}Q_{ss'}$, but
the abstract state $\hat s_k$ sends it at the single rate $\Qhat_{k\ell}$, and different
states of $B_k$ generally have different rates. The difference is the \emph{defect matrix}
$\Delta=Q\Up-\Up\Qhat\in\mathbb R^{\Sset\times\Shat}$, with entries
$\Delta_{s\ell}=Q(s,B_\ell)-\Qhat_{k\ell}$ for $s\in B_k$.
A state with a positive defect toward $B_\ell$ moves toward $B_\ell$ faster than the abstract chain does.
The rows of $\Delta$ sum to zero, and by Lemma~\ref{lem:hm-identity}, 
$\Av\Delta=\Av Q\Up-\Av\Up\,\Qhat=\Qhat-\Qhat=0$.
So single states have a nonzero defect, but within every block the defects cancel on
average. This cancellation is what makes the weights harmonic. If all blocks are singletons, or if the lumpability condition of
Prop.~\ref{prop:structure}(c) holds, then $\Delta=0$.

\begin{exmp}
Collapsing the rows of $Q$ onto $(B,G,H)$ gives $Q(s_1,\cdot)=(-2+1,1,0)=(-1,1,0)$,
where the internal arrow $s_1\to s_2$ stays in $B$, and $Q(s_2,\cdot)=(-8,6,2)$. Subtracting
$\Qhat_{B\cdot}=(-\tfrac{12}5,2,\tfrac25)$ gives the rows
$\Delta_{s_1\cdot}=(\tfrac75,-1,-\tfrac25)$ and $\Delta_{s_2\cdot}=(-\tfrac{28}5,4,\tfrac85)$
of $\Delta$; the rows of $g$ and $h$ are zero.
Each row sums to zero. With the weights $(\tfrac45,\tfrac15)$ the first column averages to
$\tfrac45\cdot\tfrac75-\tfrac15\cdot\tfrac{28}5=0$. With uniform
weights, the first column would average to $\tfrac12(\tfrac75-\tfrac{28}5)=-\tfrac{21}{10}\ne0$.
\end{exmp}

\begin{figure}[t]
 \centering
 \begin{tikzpicture}[x=0.85cm,y=0.75cm,>=Stealth]
 \draw[->,semithick] (0,0) -- (0,3.25) node[above,font=\scriptsize] {probability};
 \draw[semithick] (0,0) -- (5.0,0);
 \node[font=\scriptsize,below] at (2.95,-0.05) {the concrete states of $B_k$};
 \draw[hmgray,dashed,semithick] (0.35,1.0) -- (4.95,1.0)
   node[right,font=\scriptsize,hmgray] {$\Phat_t(k)$};
 \draw[hmblue,semithick] (0.35,1.65) -- (4.95,1.65)
   node[right,font=\scriptsize,hmblue] {$(\Av P_t)_k$};
 \foreach \x/\h in {1.6/0.75,2.5/1.25,3.4/2.05,4.3/2.55}{
   \draw[hmorange,semithick] (\x,1.65) -- (\x,\h);
   \fill (\x,\h) circle (2pt);
 }
 \node[font=\scriptsize,above right=-1pt] at (4.3,2.55) {$P_t(s)$};
 \node[font=\scriptsize,hmorange,right] at (4.3,2.1) {$y_t(s)$};
 \draw[<->,hmblue,semithick] (0.7,1.0) -- (0.7,1.65);
 \node[font=\scriptsize,hmblue,right] at (0.7,1.32) {$\varepsilon_t(k)$};
\end{tikzpicture}
 \caption{Error on one block at a fixed time. The abstraction assigns the
 single probability $\Phat_t(k)$ (dashed). The concrete probabilities $P_t(s)$ are the
 dots, and their weighted mean is $(\Av P_t)_k$ (blue). The mean error
 $\varepsilon_t(k)$ is the distance between the two lines; the spread
 $y_t(s)$ is the distance from a dot to the blue line. Thus
 $P_t(s)-\Phat_t(k)=\varepsilon_t(k)+y_t(s)$.}
 \label{fig:decomp}
\end{figure}
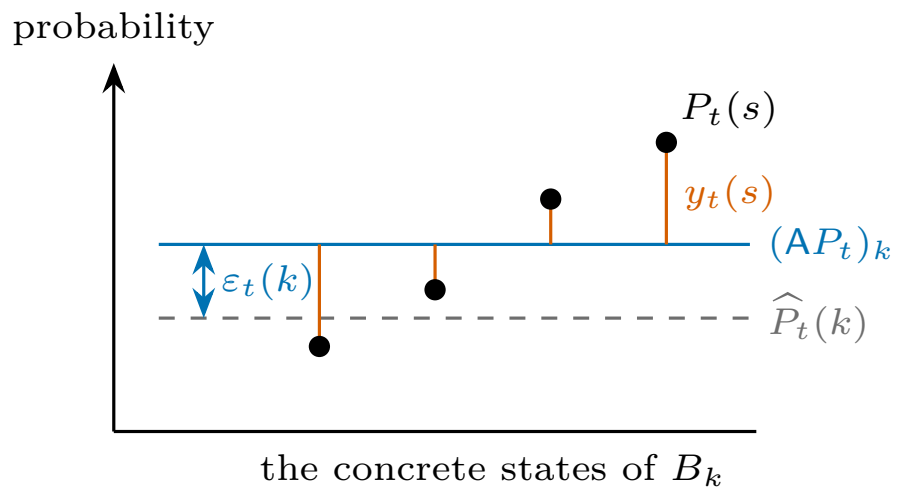

For a fixed time $t$ and block $B_k$, the abstraction assigns one
probability $\Phat_t(k)$ to every state of $B_k$, while the concrete probabilities $P_t(s)$
differ from it and from each other (Fig.~\ref{fig:decomp}). Let
$z_t=P_t-\Up\Phat_t$ be the error vector. It has two natural components: how far the
\emph{average} of the block is from the abstract value, which we call the \emph{mean error}
$\varepsilon_t$, and how far the individual states are from their own block average, which we
call the \emph{spread} $y_t$:
$\varepsilon_t=\Av z_t\in\mathbb R^{\Shat}$ and
$y_t=z_t-\Up\varepsilon_t\in\mathbb R^{\Sset}$, so that $z_t=\Up\varepsilon_t+y_t$.
By construction, $\Av y_t=0$: on every block, $y_t$ has weighted mean zero. The error we
want is $z_T(s_0)=\varepsilon_T(k_0)+y_T(s_0)$, where $k_0=\alpha(s_0)$ is the block of the initial state $s_0$.

\begin{exmp}
Consider $P=(\tfrac78,\tfrac34,1,0)$ and $\Phat(B)=\tfrac56$. As $t\to\infty$,  $z=(\tfrac1{24},-\tfrac1{12},0,0)$. The mean error is
$\varepsilon(B)=\tfrac45\cdot\tfrac1{24}-\tfrac15\cdot\tfrac1{12}=\tfrac1{60}$, and the spread on $B$ is
$y=z-\Up\varepsilon=(\tfrac1{40},-\tfrac1{10})$. Indeed $\tfrac45\cdot\tfrac1{40}-\tfrac15\cdot\tfrac1{10}=0$.
The error at $s_1$ is $\tfrac1{24}=\tfrac1{60}+\tfrac1{40}$.
\end{exmp}

\begin{lemma}[Error equations \leanproof{https://anonymous.4open.science/r/hm-mean-9D42/lean/HMAbs/Lemmas/ErrorEq.lean}]\label{lem:error-eq}
$z_0=0$, $\varepsilon_t' =\Qhat\,\varepsilon_t+\Av Q\,y_t$, and $y_t' =\Delta(\Phat_t+\varepsilon_t)+(I-\Up\Av)\,Q\,y_t$.
\end{lemma}
Lemma~\ref{lem:error-eq} is proven in Appendix~\ref{app:proof-error-eq}.

\begin{exmp}
As $t \to\infty$, both derivatives vanish, and the equations can be
checked with the numbers above. For the equation for $\varepsilon_t'$ at $B$,
$\Qhat_{BB}\varepsilon=-\tfrac{12}5\cdot\tfrac1{60}=-\tfrac1{25}$ and
$(\Av Qy)_B=-\tfrac85\cdot\tfrac1{40}-\tfrac45\cdot(-\tfrac1{10})=\tfrac1{25}$, which sum to $0$.
For the equation for $y_t'$ at $s_2$, with $\Phat+\varepsilon=(\tfrac{17}{20},1,0)$,
$\Delta_{s_2\cdot}(\Phat+\varepsilon)=-\tfrac{28}5\cdot\tfrac{17}{20}+4=-\tfrac{19}{25}$, and
$\bigl((I-\Up\Av)Qy\bigr)(s_2)=(Qy)(s_2)-(\Av Qy)_B=\tfrac45-\tfrac1{25}=\tfrac{19}{25}$. These also sum to $0$.
\end{exmp}

The equation for $\varepsilon_t'$ in Lemma~\ref{lem:error-eq} does not contain the defect
matrix $\Delta$. The mean error only follows the abstract dynamics, $\Qhat\varepsilon_t$, and is
pushed by the spread through $\Av Q\,y_t$. The rate mismatch thus does not create mean error
directly. Indeed, differentiating $\varepsilon_t=\Av z_t$ produces the extra term
$\Av\Delta\,\Phat_t$, which vanishes by $\Av\Delta=0$, and it does so only because the
averaging uses the HM weights. With uniform weights, $\Av\Delta\ne0$ (see the example above),
so this term would drive the mean error by an amount proportional to the abstract probabilities
$\Phat_t$ themselves, even when the error is still zero.

The spread is where the defect acts. In the equation for $y_t'$, $\Delta$ is applied to
$\Phat_t+\varepsilon_t=\Av P_t$, the block means of the concrete probabilities. Because every
row of $\Delta$ sums to zero, only the \emph{differences} between blocks matter, and there are two
sources of them: the mean errors $\varepsilon_t(\ell)$ of the blocks, and the gaps
$\Phat_t(\ell)-\Phat_t(k)$ between the abstract probabilities of neighboring blocks. If the
abstract probability is the same on all blocks that $B_k$ can reach, the second source disappears.
The last term, $(I-\Up\Av)Qy_t$, is the spread acting on itself. It tends to shrink the spread,
because a transition that leaves $B_k$ puts the chain into another block, where the distance
to the mean of $B_k$ is irrelevant.

\begin{corollary}[\leanproof{https://anonymous.4open.science/r/hm-mean-9D42/lean/HMAbs/Corollaries/Exact.lean}]\label{cor:exact}
If all blocks are singletons, or if Prop.~\ref{prop:structure}(c) holds
for all blocks, then $P_t=\Up\Phat_t$ for all $t\ge0$. In that case the
HM abstraction gives the exact time-bounded reachability probabilities.
\end{corollary}
Corrolary~\ref{cor:exact} is proven in Appendix~\ref{app:proof-exact}.

\subsection{The block-level bound}\label{sec:err-bound}

The equations of Lemma~\ref{lem:error-eq} are exact, but they still involve $y_t$ at every
concrete state, which is exactly what we wanted to avoid. So we keep one number per block
for the spread, its largest absolute value $\sigma_k(t)=\max_{s\in B_k}|y_t(s)|$.
The route of this subsection has three steps. First, from the two exact equations we derive
two scalar differential inequalities, one for the mean error $|\varepsilon_t(k)|$ and one for
the spread $\sigma_k(t)$. Second, in these inequalities every influence of a spread on a
quantity becomes a coefficient times one number $\sigma_\ell(t)$, and the coefficients can be read
from the rows of $Q$. Third, the inequalities for all blocks together are controlled by one linear
system with $2M$ unknowns ($M$ is the number of blocks), whose solution dominates the mean error
and the spread, and hence the error at every state. The definitions below (centered support,
active states, and the coefficients $m$, $a$, $\lambda$) are the tools for the second step.
Table~\ref{tab:error-notation} in Appendix~\ref{app:notation} collects the symbols of this section.

Every influence in the equations of Lemma~\ref{lem:error-eq} has the form $c\cdot y$: a coefficient vector
$c$ applied to a spread vector $y$ with $\sum_xw(x)y(x)=0$ and $|y|\le\sigma$. The obvious
estimate is $|c\cdot y|\le\|c\|_1\sigma$, but it wastes information, because $y$ averages to zero.
For example, if $c$ is proportional to $w$ then $c\cdot y=0$ exactly. More generally, for any
number $\beta$ we have $c\cdot y=(c-\beta w)\cdot y$, so we may subtract a multiple of $w$ from $c$
for free. The best estimate subtracts the multiple that leaves the smallest remainder.
The following quantity is that remainder.

We also need a sharper variant for the spread equation. The plain estimate above uses only
absolute values, so it discards the sign of the entry of $c$ at the state where the maximum
spread is attained. This is a loss, because that entry describes what the state does to itself:
when it leaves its block, spread is removed, and the contribution is negative. If a state $x_0$
attains the maximum, $y(x_0)=\sigma$, we know the sign of $y(x_0)$ and can keep the entry $c(x_0)$
with its sign. This gives the \emph{pinned} version below. It is the reason why we need a second
centered support, and why the diagonal coefficients $\lambda_{kk}$ introduced later can be negative.

\begin{definition}[Centered support \leanformalized{https://anonymous.4open.science/r/hm-mean-9D42/lean/HMAbs/Defs/Kappa.lean}]\label{def:kappa}
Let $w$ be a probability vector on a finite set $X$ with $w(x)>0$, and let
$c\in\mathbb R^X$. The \emph{centered support} of $c$ is
$\cs{c}{w}=\min_{\beta\in\mathbb R}\|c-\beta w\|_1$.
For $x_0\in X$ with $w(x_0)\le\tfrac12$, the version with $x_0$ pinned is
$\csp{x_0}{c}{w}=\min_{\beta\in\mathbb R}\bigl(c(x_0)-\beta w(x_0)+\sum_{x\ne x_0}|c(x)-\beta w(x)|\bigr)$.
\end{definition}

The quantity $\cs{c}{w}$ is zero exactly when $c$ is proportional to $w$, and in that case $c$
does not contribute to the estimates below. Both objectives are convex and piecewise linear in $\beta$, with breakpoints at the
ratios $c(x)/w(x)$. The first is minimized at a $w$-weighted median of these ratios.
The second is minimized at a breakpoint as well: its slope is $-1$ for $\beta\to-\infty$
and $1-2w(x_0)\ge0$ for $\beta\to+\infty$. Both are therefore computed by sorting.

\begin{exmp}
Let $c=(-\tfrac85,-\tfrac45)$, the row of $\Av Q$ on $B$, and $w=(\tfrac45,\tfrac15)$. The
ratios $c/w$ are $-2$ and $-4$, and the $w$-weighted median is $-2$. With $\beta=-2$,
$c-\beta w=(-\tfrac85+\tfrac85,\,-\tfrac45+\tfrac25)=(0,-\tfrac25)$, so $\cs{c}{w}=\tfrac25$, against
$\|c\|_1=\tfrac{12}5$.
\end{exmp}

\begin{lemma}[Centered support \leanproof{https://anonymous.4open.science/r/hm-mean-9D42/lean/HMAbs/Lemmas/Support.lean}]\label{lem:support}
Let $w$ be a probability vector on a finite set $X$ and let
$y\in\mathbb R^X$ with $\max_x|y(x)|\le\sigma$ and $\sum_xw(x)y(x)=0$.
Then for every $c\in\mathbb R^X$, $|c\cdot y|\le\sigma\,\cs{c}{w}$.
If moreover $x_0\in X$ has $y(x_0)=\sigma>0$, then $w(x_0)\le\tfrac12$ and
$c\cdot y\le\sigma\,\csp{x_0}{c}{w}$.
\end{lemma}
The proof is in Appendix~\ref{app:proof-support}. The second claim is why we only need ``active''
states below: a state attaining the maximum spread of its block has weight at most $1/2$.
\paragraph{The coefficients.}
Write $(\Av Q)_{ks'}=\sum_{s\in B_k}w_k(s)Q_{ss'}$, and call
$\Act_k=\{s\in B_k : w_k(s)\le\tfrac12\}$ the \emph{active} states of $B_k$
(a singleton block has none). Only these matter, because by Lemma~\ref{lem:support} a state
attaining the maximum spread of its block has weight at most $\tfrac12$: if $w_k(s)>\tfrac12$ and
$y(s)=\sigma$, the remaining states would have to average to $-w_k(s)\sigma/(1-w_k(s))$, which
exceeds $\sigma$ in absolute value.

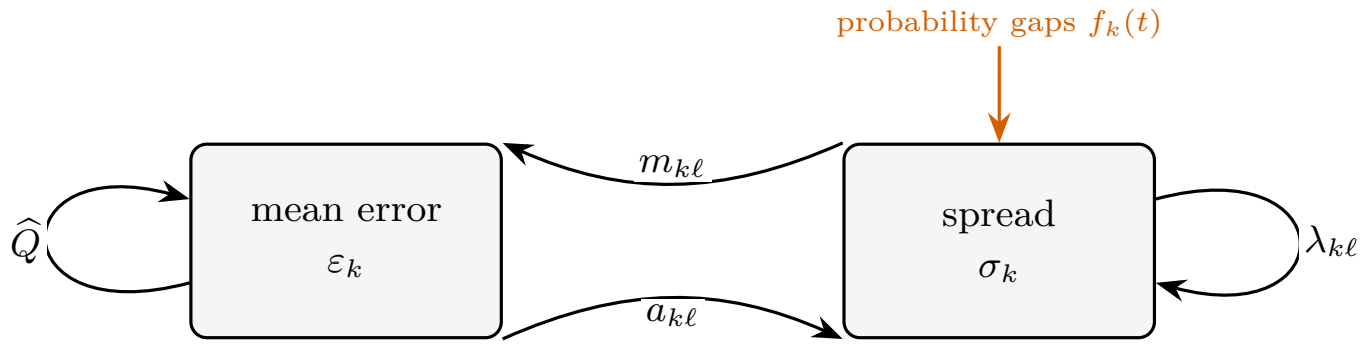
\begin{figure}[t]
 \centering
 \begin{tikzpicture}[>=Stealth,scale=0.8,transform shape]
 \node[hm box] (e) at (0,0) {mean error\\[1pt]$\varepsilon_k$};
 \node[hm box] (s) at (5.4,0) {spread\\[1pt]$\sigma_k$};
 \draw[hm arrow,bend left=24] (s.north west) to
   node[above,hm rate] {$m_{k\ell}$} (e.north east);
 \draw[hm arrow,bend left=24] (e.south east) to
   node[below,hm rate] {$a_{k\ell}$} (s.south west);
 \draw[hm arrow] (e) edge[loop left,looseness=6] node[left,hm rate] {$\Qhat$} (e);
 \draw[hm arrow] (s) edge[loop right,looseness=6] node[right,hm rate] {$\lambda_{k\ell}$} (s);
 \node[hm label,above=8mm of s,hmorange] (f) {probability gaps $f_k(t)$};
 \draw[hm arrow,hmorange] (f) -- (s.north);
\end{tikzpicture}
 \caption{Coupling of mean error and spread for a block $B_k$, with a
 neighboring block $B_\ell$. The probability gaps $f_k(t)$ (orange) are the only
 input that is not itself an error component. The coefficients are those
 defined in Sect.~\ref{sec:err-bound}.}
 \label{fig:loop}
\end{figure}

We need three families of coefficients, one for each coupling in Fig.~\ref{fig:loop}:
$m_{k\ell}$ controls the influence of the spread of $B_\ell$ on the mean error of $B_k$;
$a_{k\ell}$ controls the influence of the mean error of $B_\ell$ on the spread of $B_k$;
and $\lambda_{k\ell}$ controls the influence of the spread of $B_\ell$ on the spread of $B_k$.
Precisely, for blocks $k,\ell$ put
\begin{align*}
 m_{k\ell}&=\cs{(\Av Q)_{k\cdot}|_{B_\ell}}{w_\ell},\\
 a_{k\ell}&=\max_{s\in\Act_k}|\Delta_{s\ell}|,\\
 \lambda_{k\ell}&=\max_{s\in\Act_k}\cs{(Q_{s\cdot}-(\Av Q)_{k\cdot})|_{B_\ell}}{w_\ell}
 \quad(\ell\ne k),\\
 \lambda_{kk}&=\max_{s\in\Act_k}\csp{s}{(Q_{s\cdot}-(\Av Q)_{k\cdot})|_{B_k}}{w_k},
\end{align*}
where $|_{B_\ell}$ retains the columns indexed by $B_\ell$, and a maximum
over the empty set is $0$. \emph{All three families are read directly from the rows of $Q$, block by
block (one pass over the nonzero entries and a sort in each block), and none of them depends on
$P_t$; no concrete reachability probability is needed.} We assemble them into the
block matrix
\[
 \Gm=\begin{pmatrix}\Qhat&m\\a&\lambda\end{pmatrix}\in\mathbb R^{2M\times2M},
\]
where $m=(m_{k\ell})$, $a=(a_{k\ell})$, and $\lambda=(\lambda_{k\ell})$ are the
$M\times M$ matrices of the coefficients above. In particular, $\lambda_{kk}$ is a diagonal entry of
the lower-right block $\lambda$, not of $\Qhat$, which keeps its own diagonal entries $\Qhat_{kk}\le0$.
All off-diagonal entries of $\Gm$ are nonnegative: those of $\Qhat$ are rates, and
$m_{k\ell}$, $a_{k\ell}$, and $\lambda_{k\ell}$ ($\ell\ne k$) are maxima of $1$-norms. Only diagonal
entries can be negative: those of $\Qhat$, and the $\lambda_{kk}$, which usually are
because leaving the block removes spread. Such a matrix is called \emph{Metzler}~\cite{BermanPlemmons1994}.
This is why we assemble one matrix: the inequalities for $|\varepsilon_t|$ and $\sigma_t$
together read $\mu'\le\Gm\mu+(0,f)$ for $\mu=(|\varepsilon_t|,\sigma_t)$, and since
$e^{s\Gm}\ge0$ entrywise for $s\ge0$ when $\Gm$ is Metzler, multiplying by $e^{(t-u)\Gm}$ and
integrating preserves the inequality. So $\mu$ stays below the solution of the corresponding equation.

\begin{exmp}
Since $w_B(s_1)=\tfrac45>\tfrac12$, the only active state of $B$ is $s_2$.
Then $m_{BB}=\cs{c}{w}=\tfrac25$ by the previous example. The coefficients $a_{B\ell}$ are the absolute
entries of the row of $s_2$ in $\Delta$: $a_{BB}=\tfrac{28}5$, $a_{BG}=4$, $a_{BH}=\tfrac85$. For
$\lambda_{BB}$, subtract the averaged row from the row of $s_2$ on the columns of $B$:
$(0,-8)-(-\tfrac85,-\tfrac45)=(\tfrac85,-\tfrac{36}5)$. The choice $\beta=2$ cancels the first entry, and
with $s_2$ pinned, $\lambda_{BB}=-\tfrac{36}5-2\cdot\tfrac15=-\tfrac{38}5$. All other coefficients vanish,
because $G$ and $H$ are singletons. So only $B$ carries a nonzero comparison pair, with the coefficient matrix
$\begin{pmatrix}-12/5&2/5\\28/5&-38/5\end{pmatrix}$ on $(\varepsilon_B,\sigma_B)$.
\end{exmp}

The remaining input is the vector of \emph{probability gaps}, whose entries are
$f_k(t)=\sum_{\ell\ne k}a_{k\ell}\,|\Phat_t(\ell)-\Phat_t(k)|$
and which depend only on the abstract probabilities.

\begin{exmp}
Since $\Phat_t(G)=1$ and $\Phat_t(H)=0$,
$f_B(t)=4\bigl(1-\Phat_t(B)\bigr)+\tfrac85\Phat_t(B)$. It equals $4$ at $t=0$, and it equals $2$ once
$\Phat_t(B)=\tfrac56$. With $\Phat_t(B)=\tfrac56(1-e^{-12t/5})$ it is $2+2e^{-12t/5}$.
\end{exmp}

Figure~\ref{fig:loop} shows how the pieces fit together for one block $B_k$ and a
neighboring block $B_\ell$. The mean error is fed by the spread (coefficient $m_{k\ell}$) and otherwise
evolves under $\Qhat$. The spread is fed by the mean error (coefficient $a_{k\ell}$), by the probability
gaps $f_k$, and by itself (coefficient $\lambda_{k\ell}$). The diagonal coefficient $\lambda_{kk}$ is 
negative, expressing that the spread is lost when the chain leaves the block.

We can now state the main result: the linear system with exactly these coefficients
dominates the errors.

\begin{theorem}[Block-level error envelope \leanproof{https://anonymous.4open.science/r/hm-mean-9D42/lean/HMAbs/Theorems/Envelope.lean}]\label{thm:envelope}
Let $(\varepsilon^+,\sigma^+)\in\mathbb R^{M}\times\mathbb R^{M}$ solve
\[
 \begin{pmatrix}\varepsilon^{+}\\\sigma^{+}\end{pmatrix}'
 =\Gm\begin{pmatrix}\varepsilon^{+}\\\sigma^{+}\end{pmatrix}
 +\begin{pmatrix}0\\f(t)\end{pmatrix},
 \qquad \varepsilon^+(0)=\sigma^+(0)=0.
\]
Then $|\varepsilon_t(k)|\le\varepsilon^+_k(t)$ and $\sigma_k(t)\le\sigma^+_k(t)$ for all
$k$ and $t\ge0$. Consequently, for $s_0\in B_{k_0}$,
\[
 \bigl|P_T(s_0)-\Phat_T(\hat s_0)\bigr|
 \le\varepsilon^+_{k_0}(T)+\rho_0\,\sigma^+_{k_0}(T),
 \qquad
 \rho_0=\min\Bigl\{1,\frac{1-w_{k_0}(s_0)}{w_{k_0}(s_0)}\Bigr\}.
\]
\end{theorem}
The proof is in Appendix~\ref{app:proof-envelope}.
Here $\varepsilon^+_k(t)$ and $\sigma^+_k(t)$ are \emph{upper bounds} on the size of the mean error
$|\varepsilon_t(k)|$ and on the spread $\sigma_k(t)$ of block $B_k$; the superscript $+$ marks a
computed bound, not a true error. The final inequality says that the error at $s_0$ is at most the
mean-error bound of its block plus a share $\rho_0$ of the spread bound, where $\rho_0$ is $0$ on a
singleton block and decreases as $s_0$ gets heavier.

\paragraph{Proof idea.}
Lemmas~\ref{lem:error-eq} and~\ref{lem:support} turn the exact equations into the differential
inequality $\mu'\le\Gm\mu+(0,f)$ for $\mu=(|\varepsilon_t|,\sigma)$ (steps one and two of the route
above), and the Metzler comparison gives $\mu\le(\varepsilon^+,\sigma^+)$. The bound at $s_0$ follows
from $z_T(s_0)=\varepsilon_T(k_0)+y_T(s_0)$ and $\sum_sw_{k_0}(s)y_T(s)=0$.

\subsection{Algorithm sketch: evaluating the bound}\label{sec:err-eval}

The example above was solved in closed form because it is small. In general, the bound is
evaluated in four steps: form $\Qhat$ and the weights $w_k$ as in Sect.~\ref{sec:hm};
compute the coefficients $m_{k\ell},a_{k\ell}$ and $\lambda_{k\ell}$ of Sect.~\ref{sec:err-bound} from the rows of $Q$;
compute the abstract time-bounded probability $\Phat_t$ and the gaps $f_k(t)$ of Sect.~\ref{sec:err-bound};
and solve the $2M$-dimensional comparison system, then read off the bound of
Theorem~\ref{thm:envelope}.

Because $f_k(t)$ changes with time, the
system is not a constant-coefficient ODE. We handle this by cutting $[0,T]$ into slabs
and making $f$ constant on each slab. Since the goal is absorbing, each coordinate $\Phat_t(k)$
is non-decreasing (Lemma~\ref{lem:eval}(a)). On a slab $[t_j,t_{j+1}]$ the gap
$|\Phat_t(\ell)-\Phat_t(k)|$ may therefore be replaced by
$\max\{\Phat_{t_{j+1}}(\ell)-\Phat_{t_j}(k),\ 
\Phat_{t_{j+1}}(k)-\Phat_{t_j}(\ell),\ 0\}$,
which makes $f$ piecewise constant (Lemma~\ref{lem:eval}(b)). Each slab is then solved
by a single matrix exponential. At the nodes $t_j$ we may cap the solution by the trivial
bounds $|\varepsilon_t(k)|\le\max\{\Phat_t(k),1-\Phat_t(k)\}$, which is zero on a goal block, and
$\sigma_k(t)\le 1-\min_{s\in B_k}w_k(s)$ (Lemma~\ref{lem:eval}(c)). This is safe because
the comparison can be restarted from any valid upper bound (Lemma~\ref{lem:eval}(d)).
In floating-point arithmetic the comparison holds up to rounding error.

\begin{lemma}[Ingredients of the evaluation \leanproof{https://anonymous.4open.science/r/hm-mean-9D42/lean/HMAbs/Lemmas/Eval.lean}]\label{lem:eval}
Let $\Goal$ be an absorbing union of blocks.
\begin{enumerate}
\item[(a)] Each coordinate of $\Phat_t$ is nondecreasing in $t$.
\item[(b)] For $t\in[t_j,t_{j+1}]$ and blocks $k,\ell$,
 $|\Phat_t(\ell)-\Phat_t(k)|\le\max\{\Phat_{t_{j+1}}(\ell)-\Phat_{t_j}(k),\
 \Phat_{t_{j+1}}(k)-\Phat_{t_j}(\ell),\ 0\}$.
\item[(c)] $|\varepsilon_t(k)|\le\max\{\Phat_t(k),1-\Phat_t(k)\}$, with $\varepsilon_t(k)=0$
 if $B_k\subseteq\Goal$, and $\sigma_k(t)\le1-\min_{s\in B_k}w_k(s)$.
\item[(d)] Let $\mu=(|\varepsilon|,\sigma)$ and let $\bar f\ge f$ componentwise. If
 $\mu(t_j)\le\bar\mu$, then for $t\ge t_j$
 $\mu(t)\le e^{(t-t_j)\Gm}\bar\mu+\int_{t_j}^te^{(t-u)\Gm}(0,\bar f(u))\,du$.
\end{enumerate}
\end{lemma}
Lemma~\ref{lem:eval} is proven in Appendix~\ref{app:proof-eval}.

\section{Conclusion}\label{sec:conclusion}

We established a theoretical foundation for abstracting CTMCs using harmonic-mean aggregation. By defining abstract transition rates through a local experiment that matches mean waiting times and destination-block distributions, we provided a principled interpretation for rate construction where concrete states with longer residence times naturally receive greater weight. To address the inherent approximation of state aggregation, we developed a value-aware error bound for time-bounded reachability. Rather than treating all rate discrepancies equally, this bound leverages the computed abstract reachability probabilities and localized concrete coefficients to bound both block-mean error and within-block variation. Ultimately, this work provides a rigorous bridge between the local justification of abstract CTMC rates and computable, goal-directed error guarantees, reducing the need for expensive concrete state space evaluations. Future work includes tool implementation and empirical evaluation of these techniques.

\newpage
\bibliographystyle{splncs04}
\bibliography{ref/refs}

\newpage
\appendix
\section{Notation}\label{app:notation}

Table~\ref{tab:notation} lists the main symbols of Sects.~\ref{sec:preliminaries} and~\ref{sec:hm}, and
Table~\ref{tab:error-notation} those of Sect.~\ref{sec:error}.

\begin{table}[H]
 \centering\footnotesize
 \caption{Main notation: the original CTMC $\C$ and its blocks (left), the abstraction $\Chat$ (right).}
 \label{tab:notation}
 \setlength{\tabcolsep}{3pt}
 \renewcommand{\arraystretch}{1.05}
 \begin{tabular}{@{}c >{\raggedright\arraybackslash}p{2.9cm} @{\hspace{0.8em}} c >{\raggedright\arraybackslash}p{2.9cm}@{}}
 \toprule
 \multicolumn{2}{c}{Original CTMC} & \multicolumn{2}{c}{Abstraction}\\
 \cmidrule(r{0.4em}){1-2}\cmidrule(l{0.4em}){3-4}
 $\C=(\Sset,s_0,\R,L)$ & CTMC &
 $\Chat=(\Shat,\hat s_0,\Rhat,\Lhat)$ & abstract CTMC\\
 $\R(s,s')$ & transition rate &
 $\Rhat(\hat s_k,\hat s_\ell)$ & abstract rate (Def.~\ref{def:hm})\\
 $E(s)$ & total exit rate &
 $\Ehat_k$ & HM event rate (Sect.~\ref{sec:derive})\\
 $Q$ & generator (Sect.~\ref{sec:ctmc}) &
 $\hat\tau_k^{\mathrm{out}}$ & residence time in $\hat s_k$\\
 $\tau_s,\ \nu_s$ & residence time in $s$, next state after $s$ &
 $\hat\tau_k,\ \hat J_k$ & abstract event clock, destination index\\
 $\Pi,\ \pi_i$ & predicate set, predicate over $x_i$ (Def.~\ref{def:predicate}) &
 $\hat s_k$ & abstract state of $B_k$\\
 $B_k,\ N_k$ & block, its size (Def.~\ref{def:blocks}) &
 $\Rhat,\ \Qhat$ & abstract rates, generator\\
 $M$ & number of blocks &
 $\Eout_k$ & abstract exit rate (Sect.~\ref{sec:abstract})\\
 $\rkl{k}{\ell}{s}$ & block-level rate (Def.~\ref{def:block-rate}) &
 $\phat_{k\ell}$ & averaged branching (Sect.~\ref{sec:derive})\\
 $Z_k,\ \tau_k,\ J_k$ & local experiment: state, wait, destination (Def.~\ref{def:local}) &
 $\alpha$ & abstraction map (Def.~\ref{def:abstract-states})\\
 \bottomrule
 \end{tabular}
\end{table}

\begin{table}[H]
 \centering\footnotesize
 \caption{Symbols of Sect.~\ref{sec:error}: property and operators, error decomposition, block-level bound.}
 \label{tab:error-notation}
 \setlength{\tabcolsep}{3pt}
 \renewcommand{\arraystretch}{1.05}
 \begin{tabular}{@{}c >{\raggedright\arraybackslash}p{3.3cm} @{\hspace{0.8em}} c >{\raggedright\arraybackslash}p{3.3cm}@{}}
 \toprule
 $\Goal,\ \Goalhat$ & goal set, abstract goal set &
 $\Delta$ & defect matrix (Sect.~\ref{sec:err-equations})\\
 $P_t,\ \Phat_t$ & reachability probabilities of $\C$, $\Chat$ &
 $z_t,\ \varepsilon_t,\ y_t$ & error, mean error, spread (Sect.~\ref{sec:err-equations})\\
 $\Up,\ \Av$ & lifting, averaging &
 $\sigma_k(t)$ & spread of $B_k$\\
 $w_k(s)$ & HM weights (Prop.~\ref{prop:structure}) &
 $\varepsilon^+_k,\ \sigma^+_k$ & bounds on $|\varepsilon_k|,\ \sigma_k$\\
 $\Act_k$ & active states of $B_k$ &
 $m_{k\ell},\ a_{k\ell},\ \lambda_{k\ell}$ & coefficients (Sect.~\ref{sec:err-bound})\\
 $\kappa$ & centered support (Def.~\ref{def:kappa}) &
 $\Gm$ & comparison matrix (Sect.~\ref{sec:err-bound})\\
 \bottomrule
 \end{tabular}
\end{table}

\newpage
\section{Proofs}\label{app:proofs}

This appendix contains the proofs of all lemmas, propositions, and theorems, in the order
in which the statements appear in the paper.

\subsection{Proof of Lemma~\ref{lem:block-total} (block totals)}\label{app:proof-block-total}

\begin{proof}
The blocks partition $\Sset$, so
$\sum_{\ell}\rkl{k}{\ell}{s}=\sum_{s'\in\Sset}\R(s,s')=E(s)$,
using $\R(s,s)=0$. By the law of $\nu_s$,
$\PP(\alpha(\nu_s)=\hat s_\ell)=\sum_{s'\in B_\ell}\R(s,s')/E(s)$.
\end{proof}

\subsection{Proof of Lemma~\ref{lem:wd} (well-definedness)}\label{app:proof-wd}

\begin{proof}
(a) The off-diagonal entries of $\Qhat$ are nonnegative, because $\Ehat_k>0$
and $\phat_{k\ell}\ge0$, and every row sums to zero by the definition of $\Qhat$ in Sect.~\ref{sec:abstract}.
(b) By the semantics of a CTMC with generator $\Qhat$, the residence time in
$\hat s_k$ is exponential with rate $\Eout_k$ and independent of the next state,
which is $\hat s_\ell$ with probability $\Rhat(\hat s_k,\hat s_\ell)/\Eout_k$.
Since $\Eout_k=\Ehat_k(1-\phat_{kk})$ by Sect.~\ref{sec:abstract}, this
probability equals $\phat_{k\ell}/(1-\phat_{kk})$. (c) If $\phat_{kk}=1$, then
$\Eout_k=0$ and row $k$ of $\Qhat$ is zero. If $\phat_{kk}=0$, then
$\Eout_k=\Ehat_k$ by the decomposition of $\Ehat_k$ in Sect.~\ref{sec:abstract}.
\end{proof}

\subsection{Proof of Proposition~\ref{prop:preserve} (local preservation)}\label{app:proof-preserve}

\begin{proof}
\emph{Mean time.} By Steps~1 and~2 of Sect.~\ref{sec:derive},
$\EE[\tau_k]=\frac1{N_k}\sum_{s\in B_k}\frac1{E(s)}=\frac1{\Ehat_k}
=\EE[\hat\tau_k]$. For the second claim, a clock $\Exp(\lambda)$ has mean
$1/\lambda$. The condition $1/\lambda=\EE[\tau_k]=1/\Ehat_k$ then gives
$\lambda=\Ehat_k$.

\emph{Destination.} Conditioning on the selected concrete state gives
\[
 \PP(J_k=\ell)
 =\frac1{N_k}\sum_{s\in B_k}\PP(\nu_s\in B_\ell)
 =\frac1{N_k}\sum_{s\in B_k}\frac{\rkl{k}{\ell}{s}}{E(s)}
 =\phat_{k\ell},
\]
which includes $\ell=k$.
\end{proof}

\subsection{Proof of Proposition~\ref{prop:structure} (structure of the abstract rates)}\label{app:proof-structure}

\begin{proof}
(a) Using $\Ehat_k=N_k\big/\sum_{s'}E(s')^{-1}$,
\[
 \Ehat_k\phat_{k\ell}
 =\frac{N_k}{\sum_{s'}E(s')^{-1}}\cdot\frac1{N_k}
 \sum_{s\in B_k}\frac{\rkl{k}{\ell}{s}}{E(s)}
 =\sum_{s\in B_k}w_k(s)\,\rkl{k}{\ell}{s}.
\]
Summing over all $\ell$ and using Lemma~\ref{lem:block-total} gives the
formula for $\Ehat_k$; summing over $\ell\ne k$ gives the formula for
$\Eout_k$.

(b) If $B_k=\{s\}$, then $w_k(s)=1$, and (a) gives
$\Rhat(\hat s_k,\hat s_\ell)=\rkl{k}{\ell}{s}$. Now let all blocks be singletons.
Then $\alpha$ is a bijection. By the definition of $\hat s_0$ and $\Lhat$ in Sect.~\ref{sec:predicates}, $\alpha(s_0)=\hat s_0$ and
$\Lhat(\alpha(s))=L(s)$. For $s\ne s'$ with $B_k=\{s\}$ and $B_\ell=\{s'\}$, we have
$\rkl{k}{\ell}{s}=\R(s,s')$, so $\Rhat(\alpha(s),\alpha(s'))=\R(s,s')$ by the
first part. Both $\R$ and $\Rhat$ have no self-loops, and absorbing states
have zero rows on both sides. Hence $\Chat$ is $\C$ with its states renamed by
$\alpha$. Equality of the rates is the only non-trivial component; the state
sets, initial states and labels correspond by definition.

(c) If $\rkl{k}{\ell}{s}=c_{k\ell}$ for all $s\in B_k$, then (a) gives
\[
 \Rhat(\hat s_k,\hat s_\ell)=c_{k\ell}\sum_{s\in B_k}w_k(s)=c_{k\ell}.
\]
\end{proof}

\subsection{Proof of Lemma~\ref{lem:hm-identity} (HM identity)}\label{app:proof-hm-identity}

\begin{proof}
By Convention~\ref{conv:blocks}, every block $B_k$ either has positive exit rates
or is absorbing, and we treat the absorbing case at the end. So let $B_k$ have positive
exit rates, and write $Q(s,B_\ell)=\sum_{s'\in B_\ell}Q_{ss'}$ for $s\in B_k$. For
$\ell\ne k$, this is $\rkl{k}{\ell}{s}$, so
Prop.~\ref{prop:structure}(a) gives
$(\Av Q\Up)_{k\ell}=\sum_{s\in B_k}w_k(s)\rkl{k}{\ell}{s}=\Rhat(\hat s_k,\hat s_\ell)$.
For $\ell=k$ we have $Q(s,B_k)=\rkl{k}{k}{s}-E(s)$. Prop.~\ref{prop:structure}(a), which holds for
all $\ell$, gives
$\sum_{s\in B_k}w_k(s)\rkl{k}{k}{s}=\Ehat_k\phat_{kk}$, while
$\sum_{s\in B_k}w_k(s)E(s)=\Ehat_k$. Hence
$(\Av Q\Up)_{kk}=\Ehat_k\phat_{kk}-\Ehat_k=-\Ehat_k(1-\phat_{kk})=-\Eout_k=\Qhat_{kk}$,
where $\Eout_k=\Ehat_k(1-\phat_{kk})$ uses that the $\phat_{k\ell}$ sum to one over $\ell$
(Lemma~\ref{lem:block-total}).
If $B_k$ is absorbing, both sides are zero.
\end{proof}

\subsection{Proof of Lemma~\ref{lem:error-eq} (error equations)}\label{app:proof-error-eq}

\begin{proof}
Since $\Goal$ is a union of blocks, $\ones_\Goal=\Up\ones_{\Goalhat}$, so $z_0=0$.
Differentiating gives $z_t'=QP_t-\Up\Qhat\Phat_t=Qz_t+\Delta\Phat_t$. Apply $\Av$ and use
$z_t=\Up\varepsilon_t+y_t$, Lemma~\ref{lem:hm-identity}, and $\Av\Delta=0$:
$\varepsilon_t'=\Av Q(\Up\varepsilon_t+y_t)=\Qhat\varepsilon_t+\Av Qy_t$.
For the second equation,
$y_t'=z_t'-\Up\varepsilon_t'
=Q\Up\varepsilon_t+Qy_t+\Delta\Phat_t-\Up\Qhat\varepsilon_t-\Up\Av Qy_t$, and
$Q\Up\varepsilon_t-\Up\Qhat\varepsilon_t=\Delta\varepsilon_t$.
\end{proof}

\subsection{Proof of Corollary~\ref{cor:exact} (exactness)}\label{app:proof-exact}

\begin{proof}
Under the hypothesis, $\Delta=0$ (see the definition of $\Delta$ in Sect.~\ref{sec:err-equations}), so
$z_t'=Qz_t$ and $z_0=0$ give $z_t=0$.
\end{proof}

\subsection{Proof of Lemma~\ref{lem:support} (centered support)}\label{app:proof-support}

\begin{proof}
Since $\sum_xw(x)y(x)=0$, we have $c\cdot y=(c-\beta w)\cdot y$ for every $\beta$, and
$|(c-\beta w)\cdot y|\le\sigma\|c-\beta w\|_1$. Minimizing over $\beta$ gives the first claim of Lemma~\ref{lem:support}.
For the second claim, $\sum_xw(x)y(x)=0$ gives $w(x_0)\sigma=-\sum_{x\ne x_0}w(x)y(x)\le(1-w(x_0))\sigma$,
so $w(x_0)\le\tfrac12$. Write $c\cdot y=(c-\beta w)\cdot y$ and bound the terms with $x\ne x_0$ by
$\sigma|c(x)-\beta w(x)|$; the term with $x_0$ is $\sigma(c(x_0)-\beta w(x_0))$.
Minimize over $\beta$.
\end{proof}

\subsection{Proof of Theorem~\ref{thm:envelope} (block-level error envelope)}\label{app:proof-envelope}

\begin{proof}[of Theorem~\ref{thm:envelope}]
The vectors $z_t$, $\varepsilon_t$, and $y_t$ are built from matrix exponentials, hence
analytic in $t$. The functions $|\varepsilon_t(k)|$ and $|y_t(s)|$ are therefore
Lipschitz, and so is $\sigma_k(t)=\max_{s\in B_k}|y_t(s)|$, the maximum of finitely many
such functions. All of them are differentiable almost everywhere. We derive differential
inequalities for $|\varepsilon_t(k)|$ and $\sigma_k(t)$.

\emph{Mean error.} By Lemma~\ref{lem:error-eq},
$\varepsilon_t(k)'=\sum_\ell\Qhat_{k\ell}\varepsilon_t(\ell)
+\sum_\ell\sum_{s'\in B_\ell}(\Av Q)_{ks'}y_t(s')$. On $B_\ell$, the vector $y_t$ has
weighted mean zero and absolute value at most $\sigma_\ell(t)$, so
Lemma~\ref{lem:support} bounds the second sum by $\sum_\ell m_{k\ell}\sigma_\ell(t)$.
Since $\Qhat_{k\ell}\ge0$ for $\ell\ne k$,
\[
 |\varepsilon_t(k)|'\le\Qhat_{kk}|\varepsilon_t(k)|+\sum_{\ell\ne k}\Qhat_{k\ell}|\varepsilon_t(\ell)|
 +\sum_\ell m_{k\ell}\sigma_\ell(t).
\]

\emph{Spread.} Fix $t$ with $\sigma_k(t)>0$ at which $\sigma_k$ and all $|y_t(s)|$, $s\in B_k$, are
differentiable, and choose $s\in B_k$ with $|y_t(s)|=\sigma_k(t)$. Let $\eta=\pm1$ be
the sign of $y_t(s)$ and put $\tilde y=\eta y_t$. Then $\tilde y$ has weighted mean
zero on $B_k$, $\tilde y(s)=\sigma_k(t)$, and $|\tilde y|\le\sigma_k(t)$ on $B_k$, so
$s\in\Act_k$ by Lemma~\ref{lem:support}. The function $\sigma_k-|y_\cdot(s)|$ is
nonnegative and vanishes at $t$, so it has a minimum there and its derivative
is $0$. Hence $\sigma_k(t)'=\frac{d}{dt}|y_t(s)|=\eta\,y_t'(s)$. The rows of
$\Delta$ sum to zero, so
\[
 \Bigl|\bigl(\Delta(\Phat_t+\varepsilon_t)\bigr)(s)\Bigr|
 =\Bigl|\sum_\ell\Delta_{s\ell}\varepsilon_t(\ell)
 +\sum_{\ell\ne k}\Delta_{s\ell}\bigl(\Phat_t(\ell)-\Phat_t(k)\bigr)\Bigr|
 \le\sum_\ell a_{k\ell}|\varepsilon_t(\ell)|+f_k(t),
\]
because $s\in\Act_k$. Multiply the equation for $y_t'$ in Lemma~\ref{lem:error-eq} by $\eta$. Its last term becomes
\[
 \sum_\ell\sum_{s'\in B_\ell}\bigl(Q_{ss'}-(\Av Q)_{ks'}\bigr)\tilde y(s'),
\]
and on every block $B_\ell$ the vector $\tilde y=\eta y_t$ has weighted mean zero and
absolute value at most $\sigma_\ell(t)$. For $\ell\ne k$, Lemma~\ref{lem:support} bounds the $\ell$-th summand by
$\lambda_{k\ell}\sigma_\ell(t)$. For $\ell=k$ the summand contains $\tilde y(s)=\sigma_k(t)$
itself, and the pinned version of Lemma~\ref{lem:support} bounds it by
$\lambda_{kk}\sigma_k(t)$.
Altogether
\[
 \sigma_k'\le\sum_\ell a_{k\ell}|\varepsilon_t(\ell)|+f_k(t)+\sum_\ell\lambda_{k\ell}\sigma_\ell .
\]
At almost every $t$ with $\sigma_k(t)=0$ we have $\sigma_k'=0$, and the right-hand
side is nonnegative, so the inequality holds there too.

\emph{Comparison.} Let $\mu(t)=(|\varepsilon_t|,\sigma(t))$. We have shown
$\mu'\le\Gm\mu+(0,f)$ almost everywhere, i.e.\ $\mu'=\Gm\mu+(0,f)-r$ with
$r\ge0$, and $\mu(0)=0$ because $z_0=0$. Variation of constants gives
$\mu(t)=\int_0^te^{(t-u)\Gm}\bigl((0,f(u))-r(u)\bigr)\,du\le
\int_0^te^{(t-u)\Gm}(0,f(u))\,du=(\varepsilon^+,\sigma^+)(t)$,
because $e^{s\Gm}\ge0$ entrywise.

\emph{Endpoint.} $z_T(s_0)=\varepsilon_T(k_0)+y_T(s_0)$. From $\sum_sw_{k_0}(s)y_T(s)=0$ we get
$|y_T(s_0)|\le\frac{1-w_{k_0}(s_0)}{w_{k_0}(s_0)}\sigma_{k_0}(T)$, and trivially
$|y_T(s_0)|\le\sigma_{k_0}(T)$.
\end{proof}

\subsection{Proof of Lemma~\ref{lem:eval} (ingredients of the evaluation)}\label{app:proof-eval}

\begin{proof}
(a) Put $d_t=\Phat_t'=\Qhat\Phat_t$. For $k\in\Goalhat$ the row $k$ of $\Qhat$ is zero,
so $(d_0)_k=0$. For $k\notin\Goalhat$, $(d_0)_k=\sum_{\ell\in\Goalhat}\Qhat_{k\ell}\ge0$.
Since $d_t'=\Qhat d_t$, we get $d_t=e^{t\Qhat}d_0\ge0$, because $e^{t\Qhat}\ge0$ for a generator.
(b) By (a), $\Phat_t(\ell)-\Phat_t(k)\le\Phat_{t_{j+1}}(\ell)-\Phat_{t_j}(k)$ and
$\Phat_t(k)-\Phat_t(\ell)\le\Phat_{t_{j+1}}(k)-\Phat_{t_j}(\ell)$. The absolute value is the
larger of the two differences on the left and is at least $0$.
(c) $\varepsilon_t(k)=(\Av P_t)_k-\Phat_t(k)$ and $(\Av P_t)_k\in[0,1]$ since $P_t\in[0,1]$.
On a goal block $P_t=\Phat_t=1$. Further,
$y_t(s)=\sum_{s'\in B_k}w_k(s')\bigl(P_t(s)-P_t(s')\bigr)$, so
$|y_t(s)|\le\sum_{s'\ne s}w_k(s')=1-w_k(s)$.
(d) In the proof of Theorem~\ref{thm:envelope}, $\mu'=\Gm\mu+(0,f)-r$ with $r\ge0$.
Variation of constants from $t_j$ gives
$\mu(t)=e^{(t-t_j)\Gm}\mu(t_j)+\int_{t_j}^te^{(t-u)\Gm}((0,f)-r)\,du$, and $e^{s\Gm}\ge0$.
\end{proof}

\end{document}